\documentclass[twocolumn]{aastex62}
\usepackage{physics}
\usepackage{amsmath}
\usepackage{tabu}
\usepackage{booktabs}
\usepackage{subfigure}
\usepackage{hyperref}
\usepackage{ulem}
\usepackage{enumerate}

\newcommand{\sect}{\S~}%Section
\newcommand{\app}{Appendix~}
\newcommand{\figu}{Figure~}

\newcommand{\tab}{Table~}

\newcommand{\msun}{{M_\odot}}
\newcommand{\cc}{{\rm cm}^{-3}}
\newcommand{\K}{{\rm K}}
\newcommand{\gcc}{{\rm g~cm}^{-3}}

\newcommand{\rnew}{\zeta}
\graphicspath{{./}{figures/}}

\submitjournal{ApJ}
\shorttitle{Draft}
\shortauthors{Guo et al.}

\begin{document}

% Title
\title{Hunting for wandering massive black holes}

\author{Minghao Guo}
\affiliation{Peking University, Beijing 100871, China}
\author{Kohei Inayoshi}
\affiliation{Kavli Institute for Astronomy and Astrophysics, Peking University, Beijing 100871, China}
\email{Corresponding author: inayoshi@pku.edu.cn}
\author{Tomonari Michiyama}
\affiliation{Kavli Institute for Astronomy and Astrophysics, Peking University, Beijing 100871, China}
\author{Luis C. Ho}
\affiliation{Kavli Institute for Astronomy and Astrophysics, Peking University, Beijing 100871, China}
\affiliation{Department of Astronomy, School of Physics, Peking University, Beijing 100871, China}

\begin{abstract}
We investigate low-density accretion flows onto massive black holes (BHs) with masses of $\gtrsim 10^5~\msun$ orbiting around in the outskirts 
of their host galaxies, performing three-dimensional simulations.
Those wandering BHs are populated via ejection from the galactic nuclei through multi-body 
BH interactions and gravitational wave recoils associated with galaxy and BH coalescences.
We find that when a wandering BH is fed with hot and diffuse plasma with density fluctuations,
the mass accretion rate is limited at $\sim 10-20\%$ of the canonical Bondi-Hoyle-Littleton rate 
owing to a wide distribution of inflowing angular momentum.
We further calculate radiation spectra from radiatively inefficient accretion flows onto the wandering BH using a semi-analytical two-temperature disk model and find that the predicted spectra have a peak at the millimeter band, where the Atacama Large Millimeter/submillimeter Array (ALMA) has the highest sensitivity and spatial resolution.
Millimeter observations with ALMA and future facilities such as the next generation 
Very Large Array (ngVLA) will enable us to hunt for a population of wandering BHs and push the detectable
mass limit down to $M_\bullet \simeq 2\times10^7~\msun$ for massive nearby ellipticals, e.g., M87, 
and $M_\bullet \simeq 10^5~\msun$ for the Milky Way.
This radiation spectral model, combined with numerical simulations, will be applied to give 
physical interpretations of off-nuclear BHs detected in dwarf galaxies, which may 
constrain BH seed formation scenarios.

\end{abstract}

%%%%%%%%%%%%
%	1. Introduction    %
%%%%%%%%%%%%
\section{Introduction} 
\label{sec:intro}

Supermassive black holes (SMBHs) are harbored at the nuclei of almost all massive galaxies
in the present-day universe \citep{Kormendy2013ARA&A..51..511K}.
In the bottom-up hierarchical structure formation of the $\Lambda$ cold dark matter (CDM) cosmologies,
galaxies were assembled out of smaller mass via halo and galaxy mergers.
As a natural outcome of frequent galaxy mergers, incoming massive BHs would sink toward the centers,
form binary SMBHs at the galactic nuclei, and coalesce with gravitational wave (GW) emission, 
if the BHs were to decay their orbit via dynamical processes within a Hubble time
\citep{Begelman1980Natur.287..307B,Yu2002MNRAS.331..935Y,Merritt2013CQGra..30x4005M,
Khan2016PhRvD..93d4007K,Kelley2017MNRAS.464.3131K}.
Low-frequency GW detectors (LISA, Tianqin, Taiji) and experiments (PTA) will enable us to probe 
the cosmological evolution of SMBHs in the current framework of cosmology
\citep{Sesana2008MNRAS.390..192S,Bonetti2018MNRAS.477.3910B,Bonetti2018MNRAS.477.2599B,Bonetti2019MNRAS.486.4044B,Inayoshi2018ApJ...863L..36I,Luo2016CQGra..33c5010L}

Giant elliptical galaxies, the most massive objects in the local universe, have experienced 
a large number of merger events, predominantly minor and dry (i.e., gas-poor) mergers at lower redshifts ($z<2$), 
where their star formation activities ceased \citep{Thomas2005ApJ...621..673T}.
In gas-poor environments, multi-body BH interactions
would be one plausible way
to make BHs coalesce within a short timescale.
Because of the nature of multi-body interactions, less massive objects are likely to be ejected from the core,
leaving behind more massive binaries \citep{Bonetti2018MNRAS.477.3910B, Ryu2018MNRAS.473.3410R}.
Those ejected BHs with high velocities comparable to the escape speed from the galactic cores plunge 
into diffuse hot gas in the galactic outskirts and orbit as wandering BHs
\citep{Zivancev&Ostriker2020arXiv200406083Z}.
Similarly, the anisotropic emission of GWs (or `gravitational recoil') during the final coalescence of 
two SMBHs would make the merger remnant offset from the centers of the host galaxies
\citep{Bekenstein1973ApJ...183..657B,Campanelli2005CQGra..22S.387C,Campanelli2007ApJ...659L...5C,Campanelli2007PhRvL..98w1102C,Lousto2012PhRvD..85h4015L,Fragione2020arXiv200601867F}.
Wandering BHs can also be populated by minor galaxy mergers with significantly low mass ratios ($q\ll 0.01$)
and could fail to reach the galactic center within a Hubble time owing to slow dynamical friction
\citep{Schneider2002ApJ...571...30S,Bellovary2010ApJ...721L.148B, Tremmel2018MNRAS.475.4967T}.
However, those BHs are significantly less massive compared to the population ejected via dynamical processes
from the galactic centers.

Ejected BHs, depending on the velocity, are bound within the galactic halo potential and orbit in diffuse gas 
at velocities of $v_\infty \simeq \sigma_\star$, where $\sigma_\star$ is the stellar velocity dispersion.
When a BH with a mass of $M_\bullet$ is moving in fixed medium (or a BH stays fixed in a moving medium), 
mass accretion onto the BH begins from a characteristic radius, where the negative gravitational 
energy becomes greater than the sum of the kinetic and thermal energy of the gas. 
The so-called Bondi-Hoyle-Littleton (BHL) radius is given by
\begin{equation}
R_{\rm BHL}\equiv\frac{GM_{\bullet}}{c_\infty^2+v_\infty^2},
\label{eq:BHLradius}
\end{equation}
\citep{Bondi1952spherically}, where $G$ is the gravitational constant and $c_\infty$ is the sound speed of gas incoming from infinity.
In the classical picture, the incoming laminar flow develops a bow shock in front of the BH 
and accretes to the hole from the backward direction.
However, 3D numerical simulations find that the symmetric accretion behavior is broken by 
the instability at the shock front, leading to highly turbulent flows (see a review of earlier studies 
in \citealt{Edgar2004NewAR..48..843E}).
In the presence of a density gradient in the inflowing gas, non-zero angular momentum is carried with accreting 
turbulent matter and a disk-like structure forms around the BH \citep{Xu2019MNRAS.488.5162X}.

Generally, the outskirts of massive galaxies are filled with hot and diffuse plasma with 
a density of $n_{\rm e}\simeq 0.1~\cc$ and temperature of $T\simeq 10^7~{\rm K}$ \citep[e.g.,][]{Russell2013MNRAS.432..530R}.
Since wandering BHs are likely fed with the plasma at significantly low rates, the accretion matter does not cool 
via emitting radiation, but forms a geometrically thick and hot disk.
The solution of radiatively inefficient accretion flows (RIAFs) has been found by \citet{Ichimaru1977ApJ...214..840I} and 
studied in the subsequent works by \citet{Narayan1994ADAF_ApJ...428L..13N,Narayan1995ADAF_ApJ...444..231N}.
There are several different solutions of RIAFs, depending on what physical processes transport energy 
and angular momentum: the advection-dominated accretion flow (ADAF; 
\citealt{Narayan1994ADAF_ApJ...428L..13N,Narayan1995ADAF_ApJ...444..231N}), 
the convection-dominated accretion flow (CDAF; \citealt{Narayan2000CDAF_ApJ...539..798N, 
Quataert&Gruzinov2000ApJ...539..809Q}), and the adiabatic inflow-outflow solution (ADIOS;
\citealt{Blandford&Begelman1999MNRAS.303L...1B, Blandford&Begelman2004MNRAS.349...68B}).
In addition, numerical simulations suggest that the properties of the accretion flow are affected by
the choice of the initial conditions and boundary conditions \citep[e.g.,][]{Inayoshi2018low}.
When the gas is weakly bound to the central BH and turbulent, with a wide range of specific angular momentum
as expected for mass accretion onto a wandering BH, the overall properties of accretion differ from 
those of the known solutions.

Detecting a population of wandering BHs in the outskirts of massive galaxies is a missing link in the above scenario.
Since the electromagnetic emission from such a BH population is expected to be weak, it is difficult to identify the presence of 
accreting BHs \citep[e.g.,][]{Ho2008nuclear}.
For low-luminosity active galactic nuclei (AGNs) with radiative luminosities significantly lower than 
the Eddington value of $L_{\rm Edd}$, 
the commonly used diagnostics with optical lines are not useful \citep{Schulze2010A&A...516A..87S}.
Previous studies have focused on X-rays from low-luminosity accreting BHs 
\citep[e.g.,][]{Fujita2008ApJ...685L..59F, Fujita2009ApJ...691.1050F, Zivancev&Ostriker2020arXiv200406083Z}.
However, a long exposure time (hours) is generally required to search for and detect such dim X-ray sources
even at modest distances.

Observationally, the nuclear emission from low-luminosity AGNs is produced by synchrotron radiation 
that has a peak energy between the radio and far-infrared bands \citep[e.g.,][]{Ho1999ApJ...516..672H, Ho2008nuclear}. 
The level of radio-loudness scales inversely with the AGN activities, namely, the Eddington ratio 
\citep{Ho2002ApJ...564..120H, Sikora2007ApJ...658..815S}.
That spectral feature is also seen in the nearest SMBH, Sagittarius A$^\star$, whose activity is known to 
be very quiescent at present ($L_{\rm bol}/L_{\rm Edd} \sim 10^{-8}$; \citealt{Narayan1998ApJ...492..554N}).
The radio emission is considered to be produced from the accretion flow on the nuclear BH 
and/or by relativistic jets \citep{Narayan1995Natur.374..623N, Mahadevan1997ApJ...477..585M, 
Falcke2000A&A...362..113F, Yuan2004ApJ...606..894Y, Yuan2014ARA&A..52..529Y}.
Recent magnetohydrodynamical simulations that treat electron thermodynamics and 
frequency-dependent radiation transport suggest that synchrotron radiation is dominated in spectra of 
accretion flows at rates of $\dot{M}_\bullet \ll 10^{-5}~\dot{M}_{\rm Edd}$ 
(\citealt{Ryan2017ApJ...844L..24R}; see also \citealt{Moscibrodzka2011ApJ...735....9M}), 
where the Eddington accretion rate is defined as $\dot{M}_{\rm Edd}\equiv 10~L_{\rm Edd}/c^2$.

Motivated by this background, in this paper we investigate the dynamics of low-density
accretion flows onto a moving BH and estimate the BH feeding rate, 
performing 3D hydrodynamical simulations.
We apply the simulation results to BHs wandering at the outskirts of massive galaxies filled by hot and diffuse plasma. With a semi-analytical two-temperature disk model describing RIAFs onto BHs, we estimate that the radiation spectra have a peak in the millimeter band, where the Atacama Large Millimeter/submillimeter Array (ALMA) has the highest sensitivity and spatial resolution.
Millimeter observations with the ALMA and future facilities such as the next generation VLA (ngVLA) 
\footnote{https://ngvla.nrao.edu/}
will enable us to hunt for a population of wandering BHs.

The rest of this paper is organized as follows. 
In \sect\ref{sec:method}, we describe the methodology of our numerical simulations. 
In \sect\ref{sec:result}, we show our simulation results and explain their physical properties. 
In \sect\ref{sec:discussion}, we present the radiation spectra of wandering BHs that accrete
gas at the outskirts of different types of galaxies and discuss their detectability.
We summarize our conclusions in \sect\ref{sec:sum}.

%%%%%%%%%%
%	2. Method     %
%%%%%%%%%%
\vspace{5mm}
\section{Methodology}\label{sec:method}

We solve the 3D hydrodynamical equations using the open source code PLUTO \citep{Mignone2007PLUTO}.
The basic equations are the equation of continuity,
\begin{equation}
\frac{{\rm d}\rho}{{\rm d}t}+\rho\nabla \boldsymbol{v}=0,
\end{equation}
and the equation of motion,
\begin{equation}
\rho\frac{{\rm d}\boldsymbol{v}}{{\rm d}t}=-\nabla p -\rho\nabla\Phi,
\end{equation}
where $\rho$ is the density, $\boldsymbol{v}$ is the velocity, $p$ is the gas pressure, and
the gravitational potential is set to $\Phi=-GM_{\bullet}/r$, with $r$ the distance from the central BH.
The time derivative is the Lagrangian derivative, given by ${{\rm d}}/{{\rm d}t}=\partial/\partial t + \boldsymbol{v}\cdot\nabla$.
We solve the energy equation
\begin{equation}
    \rho\frac{{\rm d }e}{{\rm d}t}= - p \nabla \cdot \boldsymbol{v} 
\end{equation}
where $e$ is the internal energy per mass. 
The equation of state of the ideal gas is assumed as $p = (\gamma - 1)\rho e$,
where the adiabatic index $\gamma = 1.6$ here.

We introduce basic dimensionless physical quantities that characterize accretion systems of a BH with a mass of $M_\bullet$
moving at a velocity of $v_\infty$. 
If radiative and mechanical feedback associated with BH feeding are negligible, mass accretion begins from the BHL radius 
(see Eq. \ref{eq:BHLradius}) and the standard expression of the accretion rate is given by 
\begin{equation}
\dot{M}_{\rm BHL}
= \frac{4\pi G^2M^2_{\bullet}\rho_\infty}{c_\infty^3(1+\mathcal{M}^2)^{3/2}},
\end{equation}
where $\mathcal{M}(=v_\infty/c_\infty)$ is the Mach number and $\rho_\infty$ is the density of the ambient gas
\citep{Shima1985MNRAS.217..367S,Ruffert1994ApJ...427..351R}. 
The accretion rate normalized by the Eddington rate is given by
\begin{align}
\dot{m}_{\rm BHL}
& \simeq  1.5  \times10^{-6}~(1+\mathcal{M}^2)^{-3/2}\\
& \times
\left(\frac{M_\bullet}{10^{7}~\msun}\right)
\left(\frac{\rho_\infty}{10^{-25}~\gcc}\right) 
\left(\frac{T}{10^7~\K}\right)^{-3/2}.\nonumber
\end{align}
Throughout this paper, we focus on accretion flows at a low rate of $\dot{m}_{\rm BHL}\ll 10^{-4}$,
where the gas adiabaticity holds without radiative cooling, and ensure that our numerical results are scale-free.

%%% Fig.1 %%%
\begin{figure}[t]
    \centering
    \includegraphics[width=0.9\linewidth]{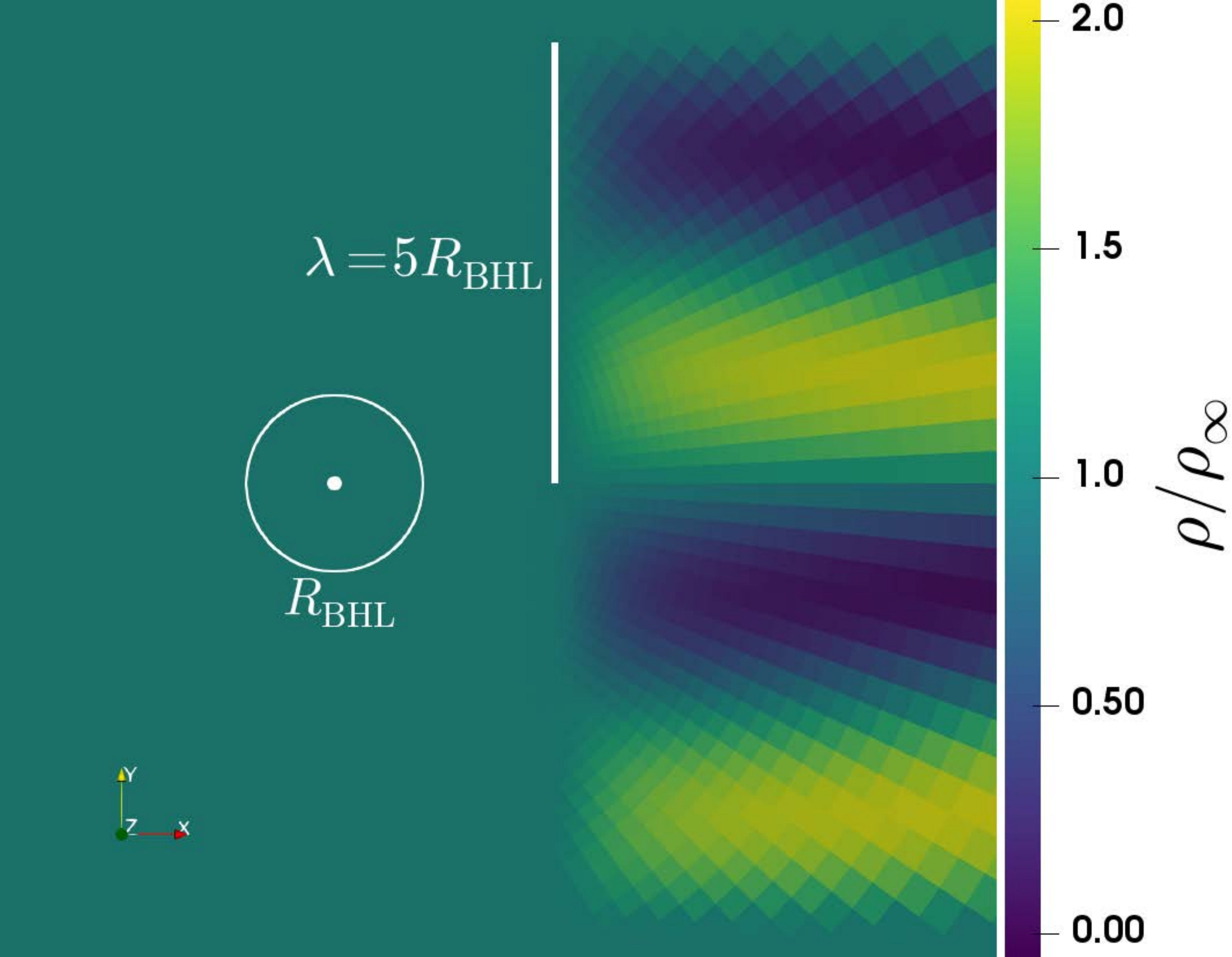}
    \caption{
    Initial density distribution on the $x-y$ plane ($\theta=\pi/2$) given by Eq. (\ref{eq:initial_rho}) with
    $\mathcal{M}=2$ and $\lambda=5R_{\rm BHL}$.
    The flow has a uniform velocity field of $\boldsymbol{v} =-v_\infty \boldsymbol{\hat{x}}$.
    }
    \label{fig:ini_rho}
    \vspace{5mm}
\end{figure}

To compute the basic equations, we employ spherical coordinates (the position of the BH is the coordinate origin)
in a three-computational domain of $r_{\rm in} \le r \le r_{\rm out}$, $\epsilon\leq \theta \leq \pi - \epsilon$ and $0\le\phi\le2\pi$, 
where $r_{\rm in }=0.08~R_{\rm BHL}, r_{\rm out }=16~R_{\rm BHL}$ in our fiducial cases and 
$\epsilon$ is set to 0.001 to avoid numerical singularity at the poles. 
We set up logarithmically spaced grids in the radial direction and uniformly spaced grids in the $\theta$- and $\phi$-directions. 
The number of grid points of our standard resolution is set to $N_r\times N_\theta \times N_\phi = 128\times 128\times 128$.
We also run simulations with a lower resolution ($N_r\times N_\theta \times N_\phi = 64\times 64\times 64$) 
and with larger values of $r_{\rm in}$, in order to check the convergence of the simulation results.

As initial conditions, we set a uniform velocity field of $\boldsymbol{v} =-v_\infty \boldsymbol{\hat{x}}$, 
where $\boldsymbol{\hat{x}}$ is the normal vector along the $x$-axis ($\theta=\pi/2,\phi=0$). 
The density distribution is given by
\begin{equation}
\label{eq:initial_rho}
    \frac{\rho}{\rho_\infty}= 
    1+A\exp{-\frac{R^2_{\rm BHL}}{(x-x_0)^2}}\sin{\frac{2\pi y}{\lambda}}\cos{\frac{2\pi z}{\lambda_0}},
\end{equation}
where the amplitude of fluctuation is set to $A=0.99$ at $x>2R_{\rm BHL}$, $|y|<\lambda$, 
and $|z|<\lambda/4$, and $A=0$ elsewhere.
The characteristic wavelengths along the $y$- and $z$-directions, which are perpendicular to the $x$-axis, 
are expressed as $\lambda$ and $\lambda_0\, (5~R_{\rm BHL})$, and we set $x_0=2~R_{\rm BHL}$.
We also impose a pressure equilibrium within the density bumps ($p_\infty=\rho_\infty c_\infty ^2/\gamma$) 
to prevent the bumpy structure from being smeared out before entering within the BH gravitational 
sphere of influence ($r<R_{\rm BHL}$). 
In our simulations, the Mach number $\mathcal{M}$ and wavelength $\lambda$ are free parameters, 
which characterize the amount of angular momentum supplied to the vicinity of the BH. 
As an example, \figu\ref{fig:ini_rho} shows the initial density distribution for the case with 
$\mathcal{M}=2$ and $\lambda=5R_{\rm BHL}$.

%%% Tab.1 %%%
\begin{table}[t]
\caption{\label{tab:para}
Parameters of the models.}
\begin{ruledtabular}
\begin{tabular} {lcccc}
Name &$\mathcal{M}$ &$\lambda(R_{\rm BHL}) $ & $r_{\rm in}(R_{\rm BHL})$ & $N_r \times N_\theta \times N_\phi$  \\
\colrule
A2    &$2$   & $5$   & 0.08 & $128^3$ \\
B2    &$2$   & $2.5$ & 0.08 & $128^3$ \\
A1    &$1$   & $5$   & 0.08 & $128^3$ \\
B1    &$1$   & $2.5$ & 0.08 & $128^3$ \\
A0    &$0.5$ & $5$   & 0.08 & $128^3$ \\
B0    &$0.5$ & $2.5$ & 0.08 & $128^3$ \\
A2r2  &$2$   & $5$   & 0.16 & $128^3$ \\
A2r4  &$2$   & $5$   & 0.32 & $128^3$ \\
A2low &$2$   & $5$   & 0.08 & $64^3$  \\
\end{tabular}
\tablecomments{Column (1): simulation name. Column (2): Mach number. Column (3): wavelength of fluctuation normalized by $R_{\rm BHL}$. 
Column (4): position of the radial inner boundary normalized by $R_{\rm BHL}$. Column (5): cell numbers.
}
\end{ruledtabular}
    \vspace{5mm}
\end{table}

%%% Fig.2 %%%
\begin{figure*}[t]
    \centering
    \includegraphics[width=\linewidth]{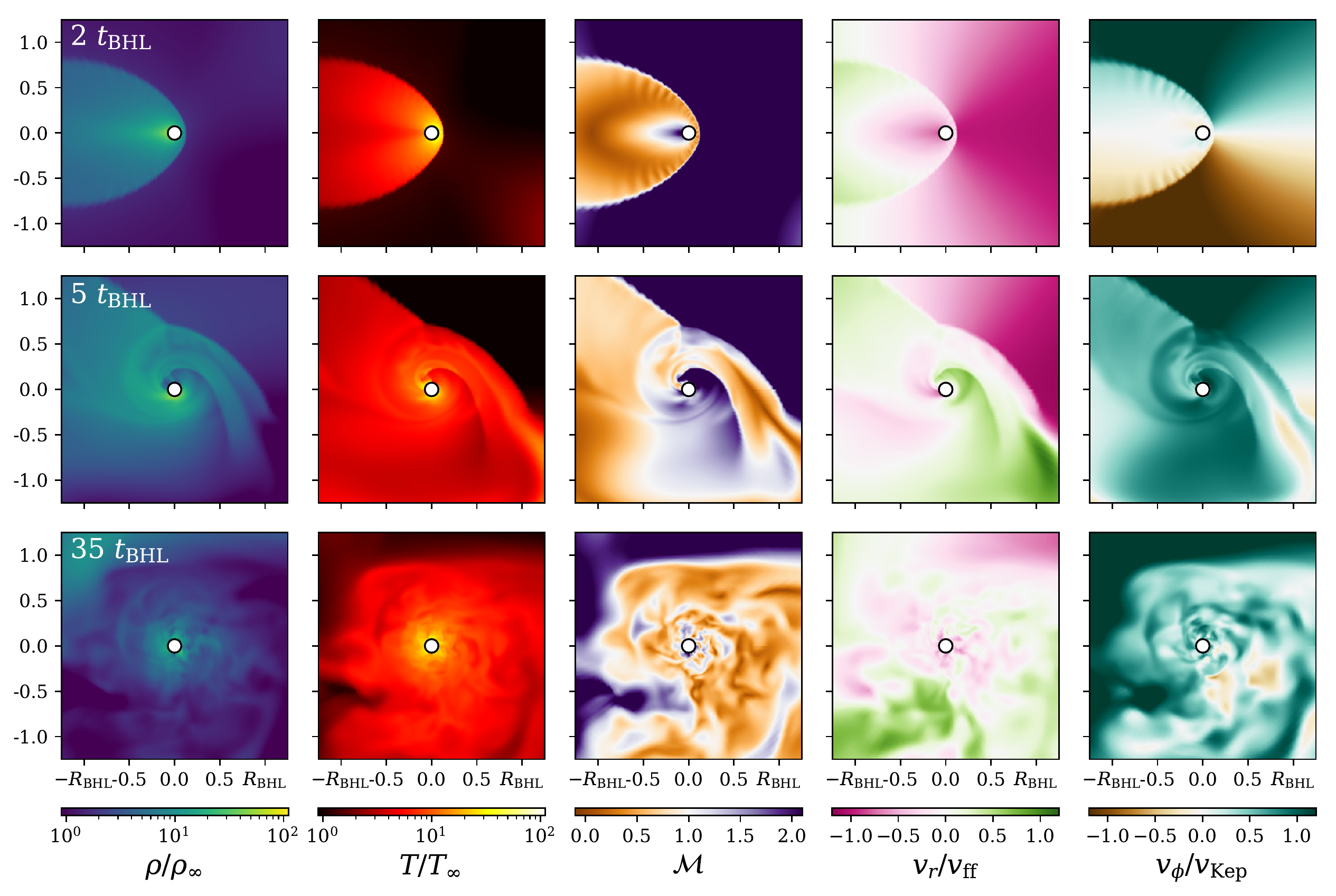}
    \caption{
    Snapshots of accretion flow for the fiducial simulation A2 at three different elapsed times of $t=2$, $5$, and $35~t_{\rm BHL}$, sliced at $z=0$. 
    From the left to the right, we show the gas density $\rho/\rho_\infty$, temperature $T/T_\infty$, Mach number $\mathcal{M}$, 
    radial velocity $v_r$ normalized by free-fall velocity $v_{\rm ff}=\sqrt{2GM_\bullet/r}$, and azimuthal velocity $v_\phi$ normalized by 
    the Keplerian velocity $v_{\rm Kep}=\sqrt{GM_\bullet/r}$. 
    The accretion flow is symmetric with respect to $y=0$ in the early stage ($t=2~t_{\rm BHL}$), forms a spiral structure owing to 
    angular momentum supply in the middle stage ($t=5~t_{\rm BHL}$), and becomes highly turbulent in the late stage ($t=35~t_{\rm BHL}$).
    } 
    \label{fig:snapshot}
    \vspace{5mm}
\end{figure*}

The outer boundary ($r=r_{\rm out}$) is divided into the upstream side ($0\leq \phi<\pi/2$ and $3\pi/2 \leq \phi \leq 2\pi$) and 
downstream side ($\pi/2\leq\phi<3\pi/2$). 
At the upstream side, we inject gas inflow at a velocity of $v_\infty$ with density set by Eq. (\ref{eq:initial_rho}).
We adopt the outflow boundary condition at the outermost grid (downstream side) and innermost grid \citep{1992ApJS...80..753S},
where zero gradients crossing the boundary are imposed on physical quantities in order to avoid spurious reflection of wave energy at the boundary.
At the inner boundary, $v_r \leq 0$ is imposed (i.e., inflowing gas from ghost cells is prohibited).
We also set a continuous condition on the poles ($\theta=0$ and $\pi$) to avoid the unphysical singularity. 
In the continuous condition, the values in the ghost cells are copied from the grids on the other side of the pole 
and the signs of $v_\theta$ and $v_\phi$ are flipped \citep{Athena2019ascl.soft12005S}.
We test that for $M_\bullet =0$ (i.e., no gravitational force) the density perturbations are advected, keeping the bumpy structure 
from the upstream ($x>0$) to the downstream ($x<0$) side without numerical diffusion and reflection due to numerical artifacts.

In \tab\ref{tab:para}, we summarize the simulation parameters we investigate in this paper.
We study the dynamics of mildly sub/supersonic gas flows ($0.5 \leq \mathcal{M} \leq 2$) because those are relevant 
to the case of BHs wandering in the outskirts of galaxies accreting hot plasma (see discussion in \S\ref{sec:discussion}).
The characteristic scale of density fluctuation $\lambda$ is set to $\sim O(R_{\rm BHL})$, in order to study the effect of 
disk formation caused by advection of angular momentum within $R_{\rm BHL}$ (in the limit of $\lambda \gg R_{\rm BHL}$, 
the flow pattern approaches the classical BHL accretion).
To see the impact of our choice of $\lambda$, we consider two cases with $\lambda = 5R_{\rm BHL}$ (run A) 
and $2.5R_{\rm BHL}$ (run B).
We also check the dependence on $r_{\rm in}$ for simulation A2. 
All the simulations last until $t=40t_{\rm BHL}$, where $t_{\rm BHL} \equiv R_{\rm BHL} / (c_\infty ^2 + v_\infty ^2)^{1/2}$
is the characteristic dynamical timescale.

%%%%%%%%%%%
%	3. Results       %
%%%%%%%%%%%

\section{Results} \label{sec:result}

\subsection{Overview of the simulations}
\label{sec:overview}

First, we discuss our fiducial case of A2, where a massive BH moves at a constant velocity with a Mach number $\mathcal{M}=2$
into hot plasma that has a density fluctuation with a characteristic wavelength $\lambda=5~R_{\rm BHL}$.
In \figu\ref{fig:snapshot}, we show the two-dimensional snapshots of the accretion flow at the plane of $z=0$ 
(i.e., perpendicular to the net angular momentum vector)
at three different elapsed times of $t/t_{\rm BHL}=2$, $5$, and $35$.
In the early stage ($t=2~t_{\rm BHL}$), the supersonic gas flow is attracted by the gravitational force of the BH and forms a bow shock 
with a symmetric structure in front of the BH.
As the density fluctuations reach within the BH influence radius ($t=5~t_{\rm BHL}$; middle panels), 
two streams both from $y>0$ and $y<0$ collide at $y=0$ and dissipate the linear momentum parallel to the $y$-axis.
Because of the density asymmetry, however, non-zero angular momentum is left behind the colliding flows, and thus 
the denser flow from $0<y/R_{\rm BHL}<0.7$ accretes onto the BH, forming spiral arms and shocks.
In the late stage after several dynamical timescales (represented at $t=35~t_{\rm BHL}$; bottom panels), 
the laminar flow with a spiral structure turns chaotic and turbulent.
Since the gas is adiabatically compressed owing to the lack of radiative cooling, thermal pressure is not negligible. 
Therefore, the turbulent flow becomes subsonic ($\mathcal{M}_{\rm tur} \approx 0.5$; third column) and the rotational velocity 
is sub-Keplerian ($v_\phi /v_{\rm Kep}\approx 0.3$; fifth column).
In this turbulent stage, the BH is fed not only through the disk but also by free-falling gas with substantially small angular momentum.

%%% Fig.3 %%%
\begin{figure*}[t]
    \centering
    \includegraphics[width=0.7\linewidth]{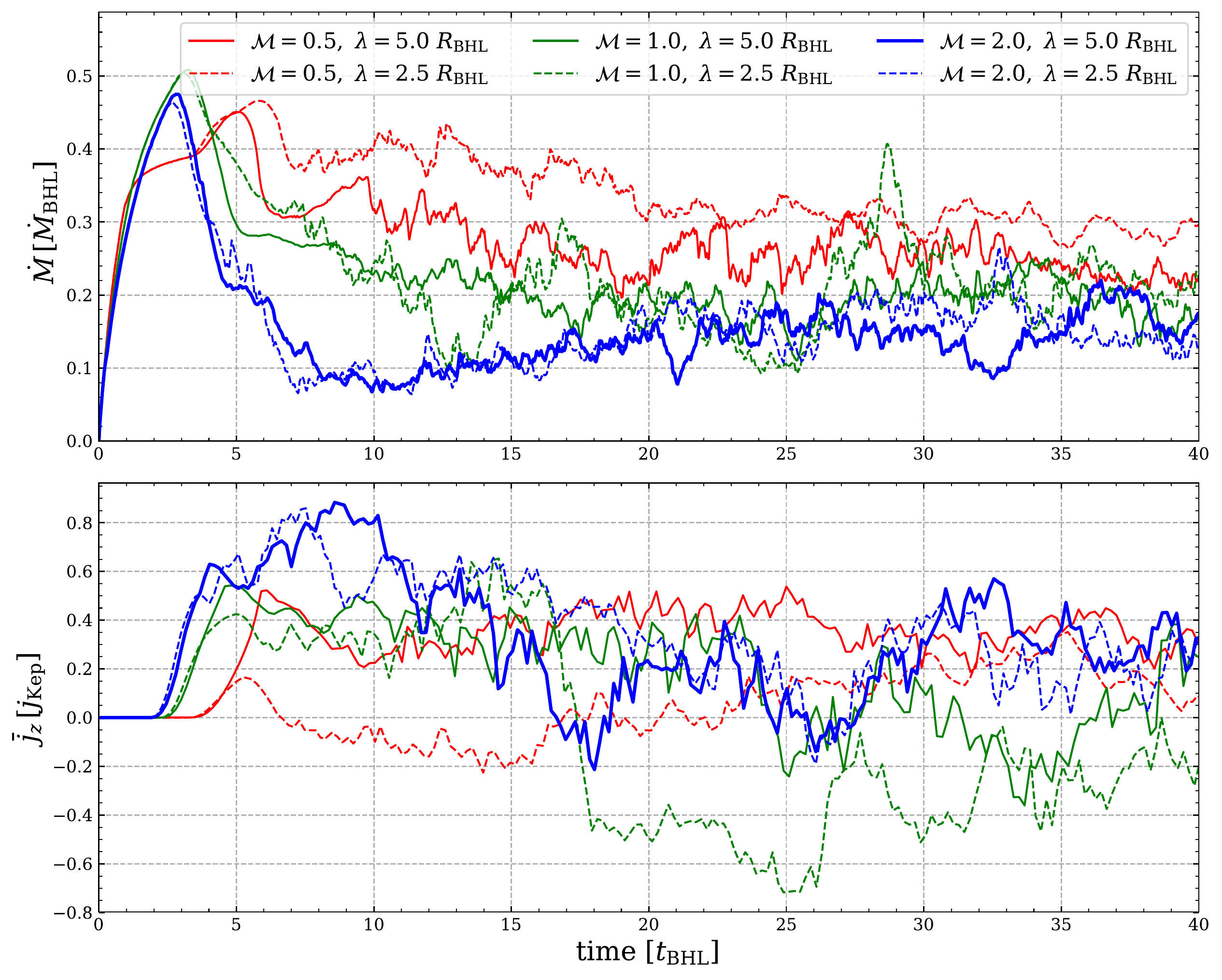}
    \caption{
    Time evolution of mass accretion rate $\dot{M}/\dot{M}_{\rm BHL}$ (top panel) and mean specific angular momentum 
    to the $z$-direction $\bar{j}_z/ j_{\rm Kep}$ of accreted materials (bottom panel) for six simulations with different values of 
    $\mathcal{M}$ and $\lambda$.
    Our fiducial model is shown by thick blue curves. 
    The accretion rates rise to $\sim 0.5~\dot{M}_{\rm BHL}$ at the beginning and drop down to $0.1-0.2~\dot{M}_{\rm BHL}$ 
    at $t>5-10~t_{\rm BHL}$, when density fluctuations enter within $R_{\rm BHL}$ and begin to form a rotating disk 
    ($\bar{j}_z \simeq 0.4~j_{\rm Kep}$).
    For all the cases, the accretion flows are in quasi-steady states. 
    }
    \label{fig:acc_rate}
    \vspace{5mm}
\end{figure*}

\figu\ref{fig:acc_rate} shows the time evolution of gas accretion rate $\dot{M}/\dot{M}_{\rm BHL}$ through 
the sink cell at $r=r_{\rm in}$ (top panel) and mean specific angular momentum to the $z$-direction 
$\bar{j}_z/ j_{\rm Kep}$ of the accreted mass (bottom panel).
Blue solid curves correspond to our fiducial case.
The accretion rate rises up to the BHL rate by $t=2~t_{\rm BHL}$ and drops to $\sim 0.1~\dot{M}_{\rm BHL}$ 
when density bumps enter within the BH influence radius and supply angular momentum of $\gtrsim 0.8~j_{\rm Kep}$ into the accreting matter.
At $t>10~t_{\rm BHL}$, mass accretion approaches a quasi-steady state at a mean rate of $\langle \dot{M} \rangle \approx 0.15~\dot{M}_{\rm BHL}$, 
though the angular momentum of the accreting matter has a large fluctuation with a mean value of $\langle \bar{j}_z \rangle /j_{\rm Kep}\approx 0.25$.

%%% Fig.4 %%%
\begin{figure*}[t]
    \centering
    \includegraphics[width=0.9\linewidth]{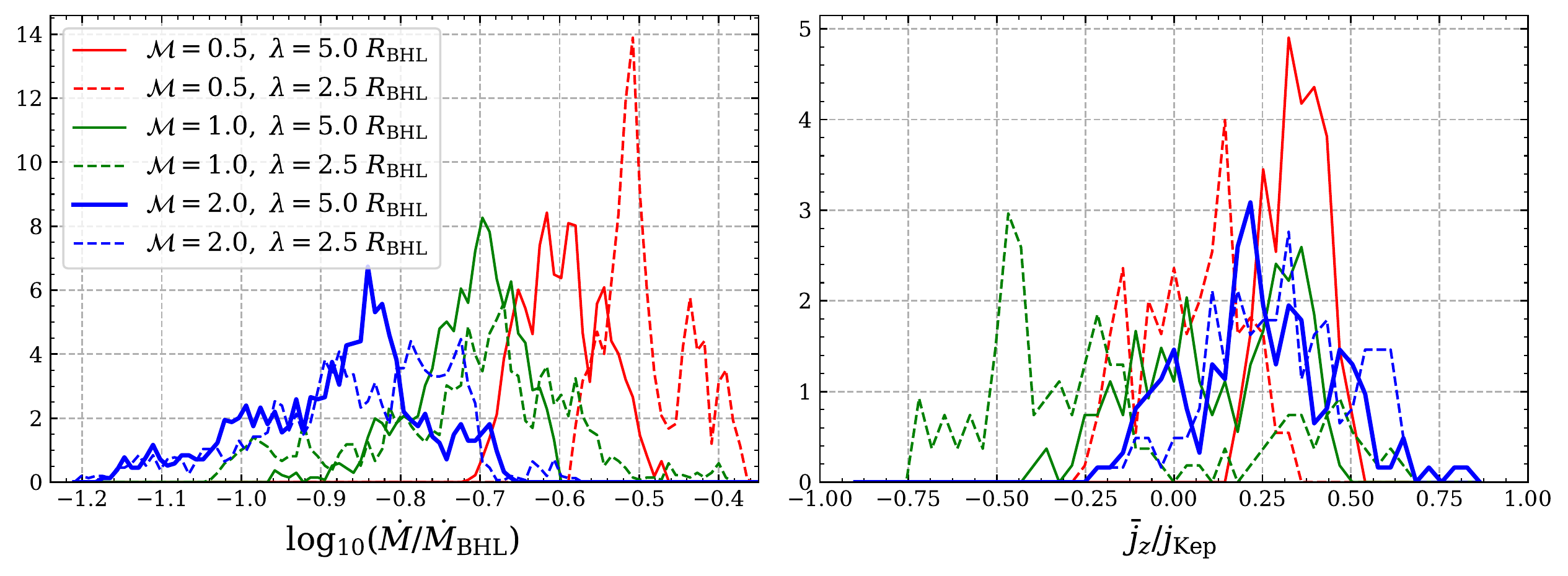}
    \caption{
    Frequency distribution of mass accretion rates $\dot{M}/\dot{M}_{\rm BHL}$ (left panel) and mean specific angular momentum 
    $\bar{j}_z/j_{\rm Kep}$ of accreted materials (right panel) during the time interval of $10\leq t/t_{\rm BHL} \leq 40$.
    Each curve corresponds to the case with $\mathcal{M}$ and $\lambda$ denoted in the left panel. 
    With higher $\mathcal{M}$, the peak value of mass accretion rate decreases and its distribution becomes wider (i.e., the flow becomes
    more unstable and turbulent).
    With smaller $\lambda$, the accretion rate and absolute values of angular momentum do not change significantly, but the sign of angular
    momentum flips due to colliding flows with different specific angular momentum (see Figure \ref{fig:snap_AB1}).
    }
    \label{fig:acc_dist}
    \vspace{5mm}
\end{figure*}

\figu\ref{fig:acc_rate} also shows the dependence of the accretion flow and its angular momentum on Mach number ($\mathcal{M}=0.5$, $1.0$, and $2.0$) 
and wavelength of the density fluctuation ($\lambda/R_{\rm BHL}=2.5$ and $5.0$), respectively.
For all the cases, the overall behavior of the accretion flow is qualitatively similar to that in our fiducial case:
the accretion rate initially increases to $\sim \dot{M}_{\rm BHL}$ and decreases to a quasi-steady value after the density bumps carry 
angular momentum within $R_{\rm BHL}$. 
In \figu\ref{fig:acc_dist}, we also calculate the frequency distribution of $\log_{10}(\dot{M}/\dot{M}_{\rm BHL})$ and $\bar{j}_z/j_{\rm Kep}$ 
during the quasi-steady state.

With a higher Mach number, the average accretion rate in the quasi-steady state tends to be lower:
$\langle \dot{M} \rangle /\dot{M}_{\rm BHL} \simeq 0.25$, $0.20$, and $0.15$ in the simulations of A0, A1, and A2, respectively.
The angular momentum of accreting matter weakly depends on the Mach number, and the peak value is kept at 
$\langle \bar{j}_z \rangle \approx 0.3~j_{\rm Kep}$.
Besides, as shown in \figu\ref{fig:acc_dist} (solid curves), the width of the distributions becomes wider as the Mach number increases.
This indicates that the accretion flow becomes more unstable and turbulent for higher values of $\mathcal{M}$.
Note that since $\dot{M}_{\rm BHL}\propto (1+\mathcal{M}^2)^{-3/2}$, the accretion rate is reduced by a factor of $\simeq 13.3$ from
the A0 run ($\mathcal{M}=0.5$) to the A2 run ($\mathcal{M}=2$).

%%% Fig. 5 %%%
\begin{figure*}[t]
    \centering
    \includegraphics[width=0.7\linewidth]{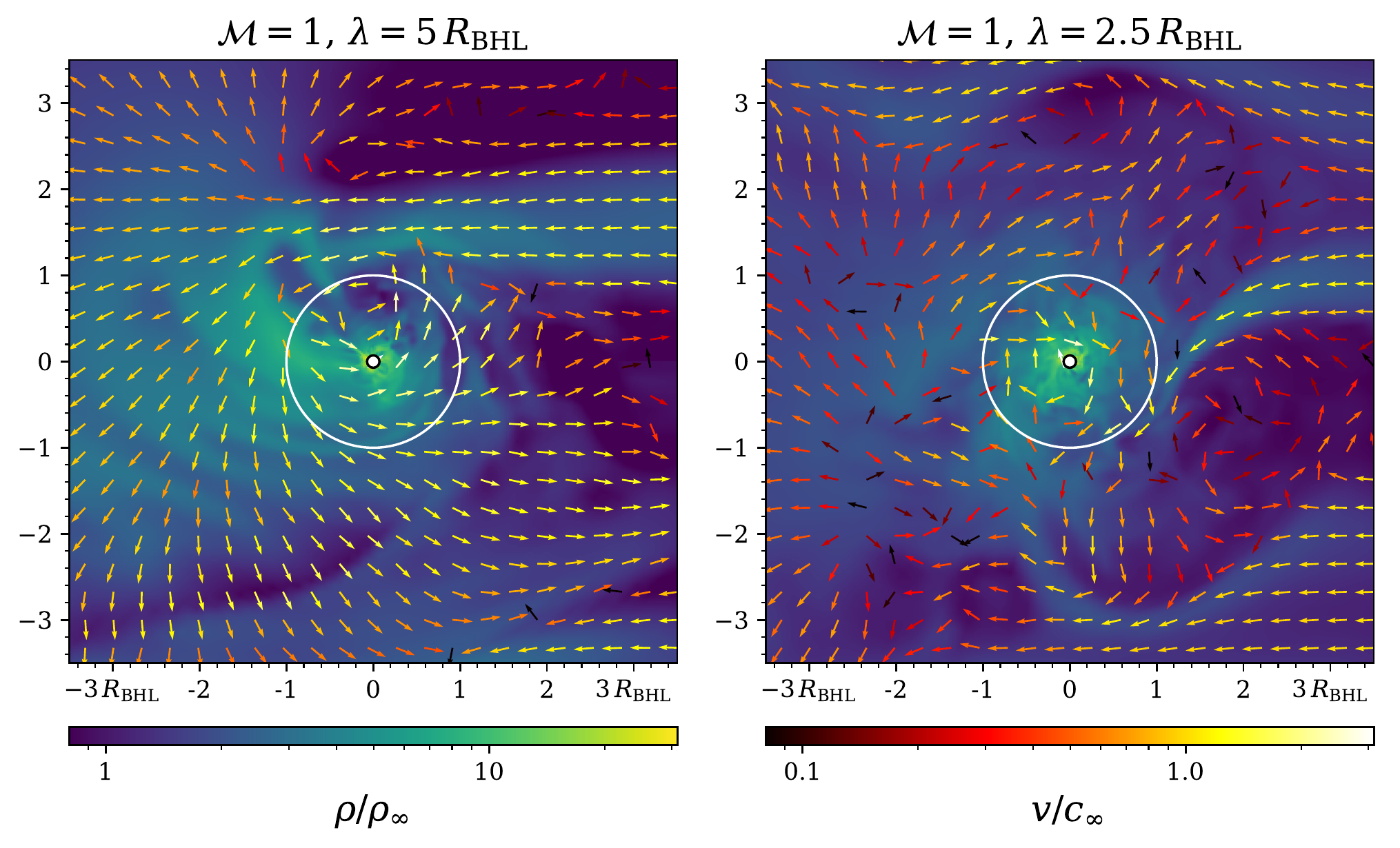}
    \caption{
    Snapshots of the density distribution and velocity vectors of accretion flow for simulations A1 (left) and B1 (right) 
    at $t=20~ t_{\rm BHL}$, sliced at $z=0$. 
    With the larger $\lambda$ (left), the angular momentum of the flow within $\sim R_{\rm BHL}$ (white circle) is dominated by 
    the incoming stream from $0\leq y \leq \lambda/2$, leading to $j_{\rm z}>0$.
    With the smaller $\lambda$ (right), the net angular momentum is determined by complex collisions of flows with different 
    angular momentum, inducing the flip of the rotating direction.
    }
    \label{fig:snap_AB1}
    \vspace{5mm}
\end{figure*}

With a shorter wavelength of density fluctuation, the flow pattern becomes more complex,
although the absolute values of accretion rates and angular momentum do not change significantly (dashed curves in Figures \ref{fig:acc_rate} and \ref{fig:acc_dist}).
In \figu\ref{fig:snap_AB1}, we show the distribution of the gas density and velocity vector at an elapsed time of $t=20~t_{\rm BHL}$ for the A1 (left) and B1 (right) runs, respectively.
When the half wavelength is sufficiently larger than $R_{\rm BHL}$ as shown in the left panel, the incoming stream from $0\leq y \leq \lambda/2$ supplies mass and 
angular momentum with $j_{\rm z}>0$ (i.e., the counterclockwise direction) within $R_{\rm BHL}$.
On the other hand, in the right panel, the incoming stream from $-\lambda/2\leq y \leq -\lambda$ carries a larger amount of angular momentum and flips 
the direction of angular momentum (see also the bottom panel of \figu\ref{fig:acc_rate} at $t=20~t_{\rm BHL}$).
As a result of the flow collisions around $\sim R_{\rm BHL}$, the accretion flow turns highly turbulent, and thus the angular momentum distribution becomes wider.

\subsection{The properties of the accretion flows}\label{sec:disk_property}
Next, we describe the properties of the accretion flow onto a moving BH, considering the time-averaged profiles of physical quantities.
In the following, we show time-averaged values over $10\leq t/t_{\rm BHL} \leq 40$.

%%% Fig.6 %%%
\begin{figure}
\centering
\includegraphics[width=\linewidth]{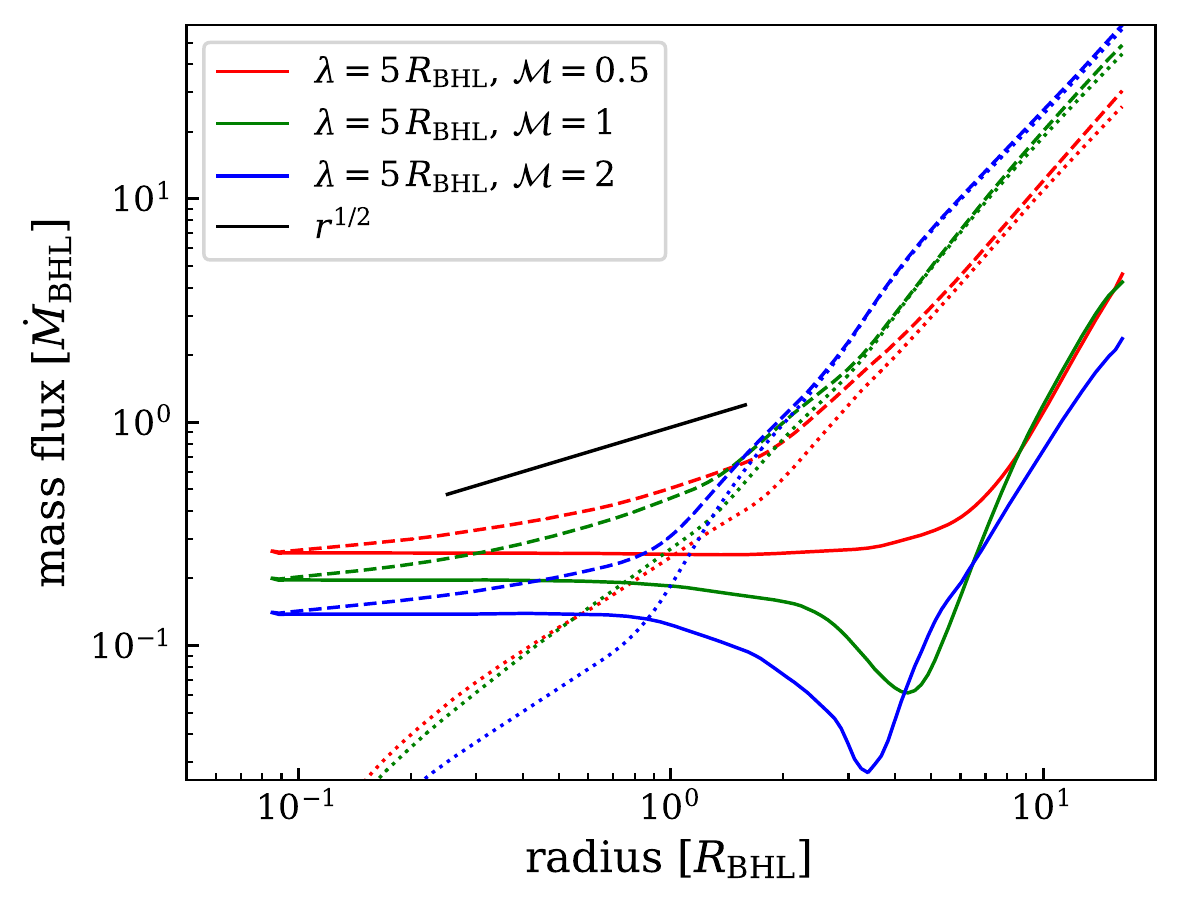}
\caption{
Time-averaged radial profiles of the mass inflow (dashed), outflow (dotted), and net (solid) accretion rate 
for the A0 (red), A1 (green), and A2 (blue) simulations.
Within the BH influence radius ($r<R_{\rm BHL}$), the inflow rate follows $\dot{M}_{\rm in} \propto r^{1/2}$
and the net accretion rate becomes a constant value.
       }
    \label{fig:massflux_Ma}
    \vspace{5mm}
\end{figure}

\figu\ref{fig:massflux_Ma} shows the radial structure of the angle-integrated mass inflow (dashed) and outflow (dotted) rates for 
the A0, A1, and A2 simulations.
These rates are defined as
\begin{equation}
    \dot{M}_{\rm in}= -\int_0^{2\pi}\int_0^{\pi} \langle \rho \cdot \min(v_r,0) \rangle r^2 \sin \theta d\theta d\phi,
\end{equation}
\begin{equation}
    \dot{M}_{\rm out}= \int_0^{2\pi}\int_0^{\pi} \langle \rho \cdot \max(v_r,0) \rangle  r^2 \sin \theta d\theta d\phi,
\end{equation}
where $\langle \cdot\cdot\cdot \rangle$ means the time-averaged value.
We also define the net accretion rate by $\dot{M} \equiv \dot{M}_{\rm in}- \dot{M}_{\rm out}$ (solid).
Note that both the inflow and outflow rates are proportional to the area ($\propto r^2$) at larger radii where 
a uniform medium moves with a constant velocity without being affected by the gravitational force of the BH\footnote{
The time-averaged values of $\dot{M}$ at larger radii do not converge to zero because the flows at $|x|\gg R_{\rm BHL}$ are not fully symmetric
within $t<40~t_{\rm BHL}$.}.
Within the BH influence radius ($r<R_{\rm BHL}$), the mass inflow rate starts to deviate from $\dot{M}_{\rm in}\propto r^2$ and approaches
$\dot{M}_{\rm in}\propto r^{1/2}$, while the outflow rate decreases toward the center.
As a result, the net accretion rate is nearly constant, and the accretion system is in a quasi-steady state.
The radial dependence of the mass inflow rate $\dot{M}_{\rm in}\propto r^{1/2}$ is consistent with the result of simulations 
where mass accretion with a broad range of angular momentum occurs 
\citep{Ressler2018MNRAS.478.3544R, Xu2019MNRAS.488.5162X}.
We note that this accretion solution is different from those of self-similar RIAF solutions for a static BH (see also discussion below): 
$\dot{M}_{\rm in}\propto r^{0}$ (ADAF; \citealt{Narayan1995ApJ...452..710N}) and $\dot{M}_{\rm in}\propto r$ (CDAF; \citealt{Quataert&Gruzinov2000ApJ...539..809Q}, \citealt{Inayoshi2018low}).

%%% Fig. 7 %%%
\begin{figure}[t]
    \centering
    \includegraphics[width=0.95\linewidth]{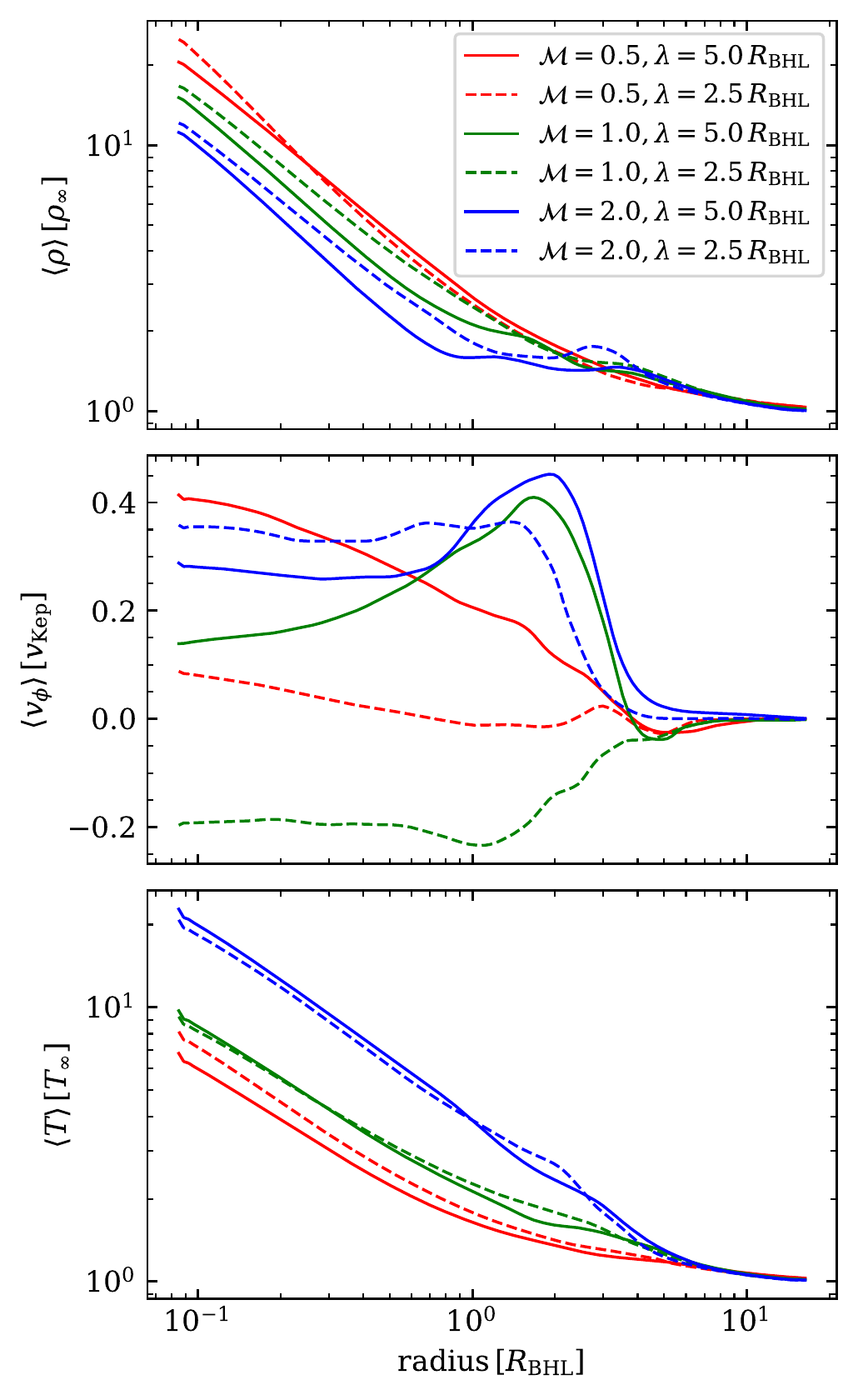}
    \caption{
Time and angle-averaged radial structures of the density (top), azimuthal velocity (middle), and temperature (bottom)
for different values of $\mathcal{M}$ and $\lambda$. 
The density and temperature profiles within $R_{\rm BHL}$ are well approximated by a power-law distribution of $\rho \propto r^{-1}$ and 
$T \propto r^{-1}$, respectively. 
    }
    \label{fig:disk_pro}
    \vspace{5mm}
\end{figure}

%%% Fig. 8 %%%
\begin{figure}[t]
    \centering
    \includegraphics[width=0.94\linewidth]{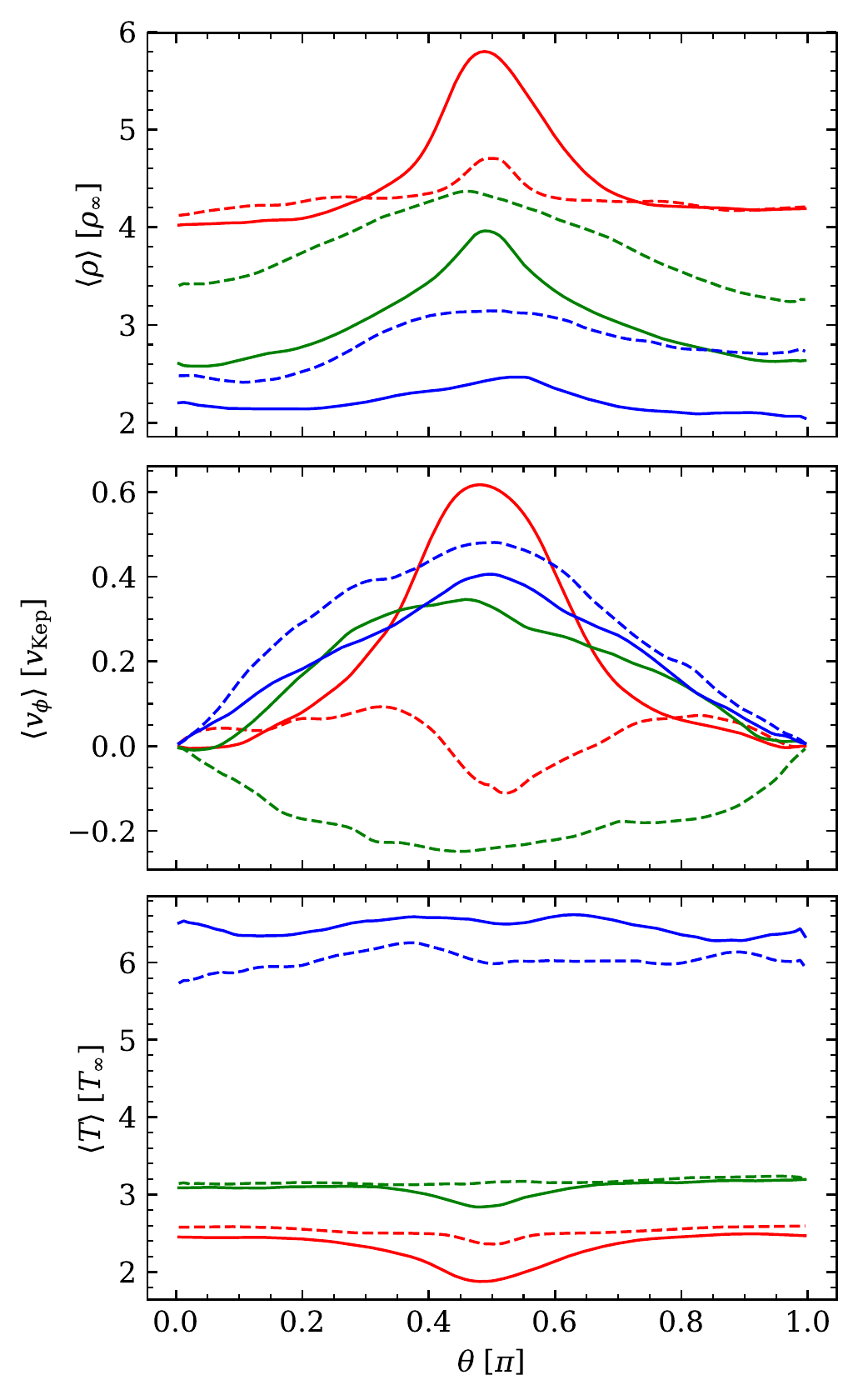}
    \caption{
    Same as \figu\ref{fig:disk_pro}, but for the angular profiles at $r=0.2~R_{\rm BHL}$. 
    }
    \label{fig:disk_pro_theta}
    \vspace{5mm}
\end{figure}

In \figu\ref{fig:disk_pro}, we present the angle-averaged radial profiles of the density, rotational velocity, and temperature for the six models.
For all the cases, the density and temperature begin to increase toward the center within the BH influence radius ($r\lesssim 2~R_{\rm BHL}$), 
where the accretion flow forms a sub-Keplerian rotating disk with a mean velocity $|v_\phi|\approx (0.2-0.4)~ v_{\rm Kep}$.
Since the flow is not fully supported by the centrifugal force, the time-averaged inflow velocity is comparable to $\sim (0.3-0.5)v_{\rm Kep}$.
As the inflow rate in the quasi-steady state is approximated as $\dot{M}_{\rm in} \propto r^{1/2}$, the density follows $\rho \propto r^{-1}$ (see the top panel of Figure \ref{fig:disk_pro}).
Since radiative cooling is neglected in our simulations, the accretion flow is adiabatically compressed by the gravity of the BH and 
the temperature increases to the center following $T\propto r^{-1}$, as expected from energy conservation. 
Note that this treatment is valid only when the BH is embedded in a low-density diffuse plasma so that the radiative cooling time is longer 
than the dynamical timescale at $r\simeq R_{\rm BHL}$ (and the orbital timescale for wandering BHs at the outskirts of galaxies; see \S\ref{sec:discussion}).
In \figu\ref{fig:disk_pro_theta}, we show the time-averaged angular profiles at $r=0.2~R_{\rm BHL}$ for the same physical quantities shown in \figu\ref{fig:disk_pro}.
Although the density and rotational velocity increase around the equatorial plane, the accretion flow is no longer a geometrically thin disk structure.

The power-law density profile ($\rho\sim r^{-1}$) is qualitatively different from those of known RIAFs:
$\rho \sim r^{-3/2}$ for ADAF solutions \citep{Narayan1995ApJ...452..710N} and $\rho\sim r^{-1/2}$ for CDAF solutions 
\citep{Quataert&Gruzinov2000ApJ...539..809Q, Inayoshi2018low}.
The overall properties of the accretion flow are similar to those discussed by 
\citet{Ressler2018MNRAS.478.3544R} and \citet{Xu2019MNRAS.488.5162X}, 
where the angular momentum of accretion flows is widely distributed.

\figu\ref{fig:disk_pro} also shows the dependence of the physical quantities on the Mach number and wavelength of density fluctuation.
While the density and temperature hardly depend on the choice of $\lambda$, the density decreases and temperature increases with higher values of $\mathcal{M}$.
The density reduction simply reflects the dependence of $\dot{M}_{\rm in}$ on the Mach number due to the input of different angular momentum within 
the BH influence radius, as shown in Figures \ref{fig:acc_rate}, \ref{fig:acc_dist}, and \ref{fig:massflux_Ma}.
We note that the $\mathcal{M}$ dependence of temperature is not true, but is caused by the radius being normalized by the BHL radius.
In adiabatic gas, the temperature is given by the virial temperature independent of $\mathcal{M}$; 
$T\propto GM/r \propto (1+\mathcal{M}^2)(r/R_{\rm BHL})^{-1}$.
The amplitude of the rotational velocity is a fraction of the Keplerian velocity within $r \lesssim 2~R_{\rm BHL}$,
though the rotation direction is more time-dependent for shorter wavelengths, as shown in \figu\ref{fig:acc_dist}.

%%% Fig.9 %%%
\begin{figure}[t]
    \centering
    \includegraphics[width=\linewidth]{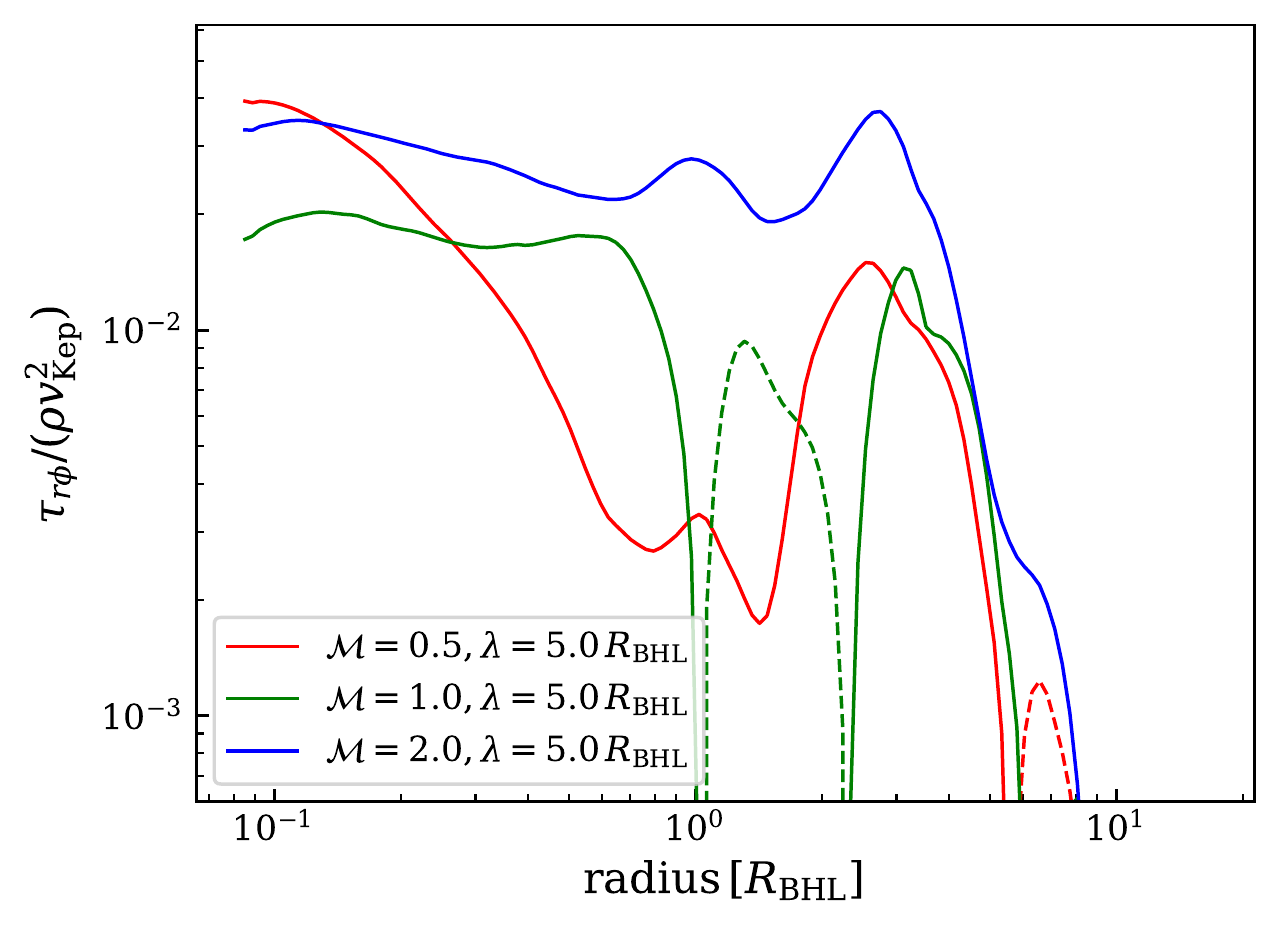}
    \caption{
    Radial profiles of the $r-\phi$ component of the Reynolds stress $\tau_{r\phi}$ normalized by $\rho v_{\rm Kep}^2$
    for the cases with different Mach numbers. 
    The solid and dashed curves show positive and negative values, respectively. 
    Since the Reynolds stress is approximated as $\tau_{r\phi}\propto Ar^{-2}$, where $A$ is positive, 
    turbulent motions transport angular momentum outward within the BH influence radius.
      }
    \label{fig:vis}
        \vspace{5mm}
\end{figure}

We note that our simulations do not treat an explicit viscosity. 
As discussed in previous studies
\citep{Igumenshchev2000ApJS..130..463I, Igumenshchev2000ApJ...537L..27I, Narayan2000CDAF_ApJ...539..798N, 
Quataert&Gruzinov2000ApJ...539..809Q, Igumenshchev2003ApJ...592.1042I},
the angular momentum of the accretion flow can be transported by turbulence excited by colliding flows.
To analyze the effect, we calculate the $r-\phi$ component of the mass-weighted Reynolds stress,
\begin{equation}
    \tau_{r\phi} = \frac{1}{4\pi} { \int\limits_0^{2\pi} \int\limits_0^{\pi} \langle \rho v'_r v'_\phi \rangle \sin{\theta}{\rm d}\theta {\rm d}\phi},
\end{equation}
where $v'_i\equiv v_i - \langle v_i \rangle$.
In \figu\ref{fig:vis}, we show the radial profile of the Reynolds stress normalized by $\rho v_{\rm Kep}^2$ for the three cases.
The Reynolds stress increases with the Mach number because the flow is more turbulent, and for $\mathcal{M}\gtrsim 1$ it is approximated by 
$\tau_{r\phi}\simeq Ar^{-2}$, where $A$ is positive.
This positive value of $\tau_{r\phi}$ indicates that the turbulent motions transport angular momentum outward.
By analogy with the standard $\alpha$-viscosity model
\citep{Shakura&Sunyaev1973A&A....24..337S},
we define the effective viscous parameter by 
\begin{equation}
    \hat{\alpha}(r) \equiv \frac{ \int\limits_0^{2\pi} \int\limits_0^{\pi} \langle \rho v'_r v'_\phi \rangle \sin{\theta}{\rm d}\theta {\rm d}\phi}
    {\int\limits_0^{2\pi} \int\limits_0^{\pi} \langle \rho c_s^2 \Omega /\Omega_{\rm Kep} \rangle \sin{\theta} {\rm d}\theta {\rm d}\phi},
\end{equation}
to quantify the strength of turbulent viscosity.
In our simulations, we obtain $\hat{\alpha} (r)\simeq 0.2$ within $r\simeq R_{ \rm BHL}$. 
Therefore, turbulence transports angular momentum effectively even without MHD effects. 
Recently, \citet{Ressler2020MNRAS.492.3272R} found that MHD and pure-HD simulations show similar properties of wind-fed accretion flows onto a BH in a nuclear region. In their situation, similarly to our simulations, mass accretion is allowed owing to a wide distribution of angular momentum provided stellar winds, even absent much angular momentum transport led by the MRI.

\subsection{Dependence on $r_{\rm in}$}
\label{sec:r_in}
\begin{figure}[t]
    \centering
    \includegraphics[width=\linewidth]{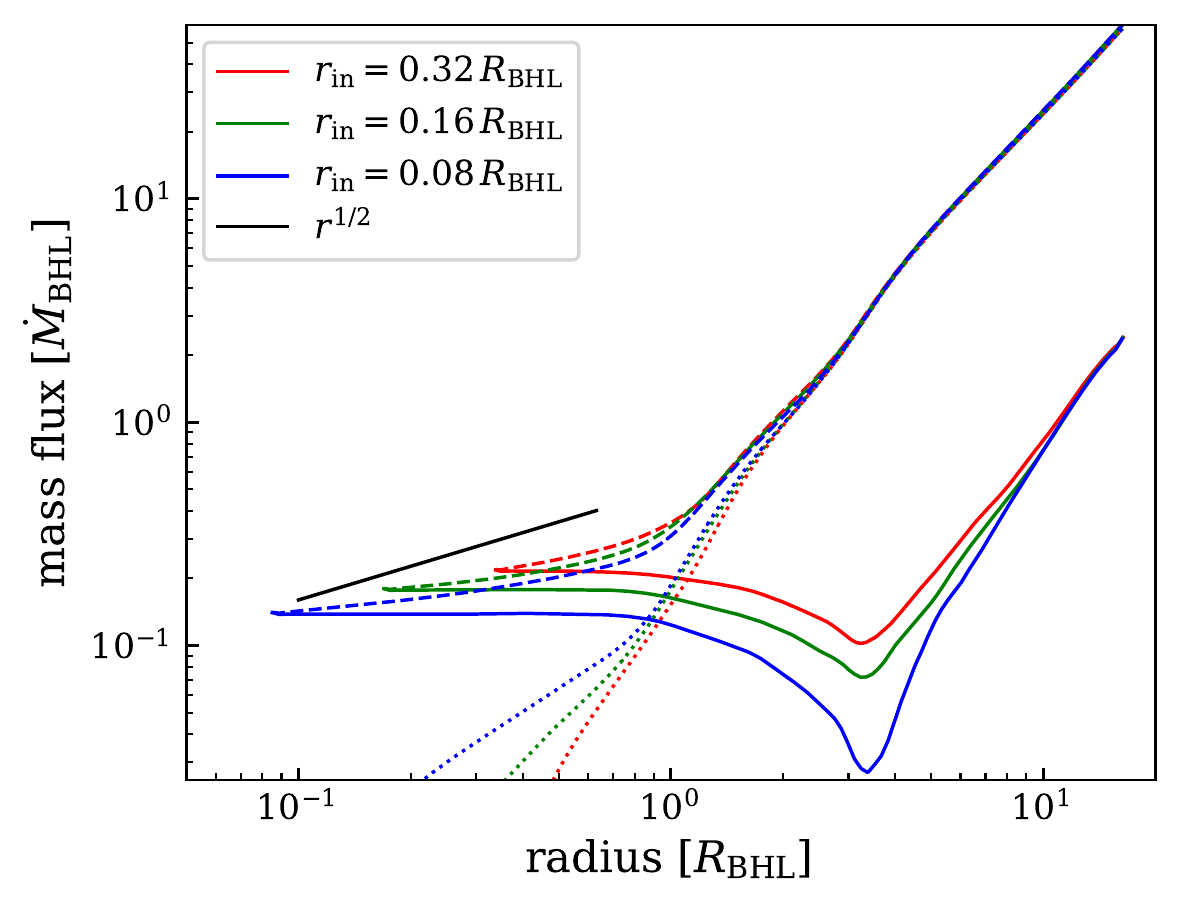}
    \caption{
    Time-averaged radial structures of the mass inflow (dashed), outflow (dotted), and net (solid) accretion rate for simulations 
    with different sizes of the innermost grid $r_{\rm in}$.
    The net accretion rate follows $\propto r_{\rm in}^{1/2}$.
    }
    \label{fig:massflux_rin}
    \vspace{5mm}
\end{figure}

Because of limitations in computing time, we do not extend our computational domain down to the BH event horizon scale ($r \sim r_{\rm Sch}$). 
Instead, we conduct two additional simulations with different locations of the innermost grid, at $r_{\rm in}/R_{\rm BHL} = 0.16$ and $0.32$.
\figu\ref{fig:massflux_rin} shows the radial profiles of time-averaged and angle-integrated mass inflow rate (dashed), outflow rate (dotted),
and net accretion rate $\dot{M}$ for each value of $r_{\rm in}$. 
Within the BH influence radius, the inflow rate dominates the outflow rate, and the net rate becomes constant for all the cases.
The normalization of the net accretion rate nicely scales with $\dot{M}_{\rm in}(r=r_{\rm in})\propto r_{\rm in}^{1/2}$.
In \app\ref{appendix:toymodel}, we describe the physical reason why the inflow rate depends on $r^{1/2}$ with an analytical model.

In order to check whether radiative cooling matters, we compare the the heating timescale to the cooling timescale. 
Since $\rho\simeq \rho_\infty(r/R_{\rm BHL})^{-1}$ and $T\simeq T_\infty (r/R_{\rm B})^{-1}$, the timescale for free-free emission at the rate of 
$Q_{\rm br}^{-}\propto \rho^2T^{1/2} \propto r^{-5/2}$ is estimated as $t_{\rm cool}\propto \rho T/Q_{\rm br}^{-} \propto r^{1/2}$. 
Since the main heating source in a RIAF is viscous dissipation, the heating timescale is given by 
$t_{\rm vis}\simeq (\gamma(\gamma-1)\hat{\alpha}\Omega)^{-1} \propto r^{3/2}$, where 
$\hat{\alpha}\simeq 0.2$ and $\Omega \simeq 0.3~\Omega_{\rm Kep}$ for our case.
Thus, the ratio of the two timescales is estimated as
\begin{equation}
\begin{split}
    \frac{t_{\rm vis}}{t_{\rm cool}}
    \simeq  &~0.14~ \left(\frac{\hat{\alpha}}{0.2}\right)^{-1} \left(\frac{r}{R_{\rm BHL}}\right)
    \left(\frac{\dot{m}_{\rm B}}{10^{-3}}\right)\\
    &\times \left(\frac{T_\infty}{10^7~{\rm K}}\right)^{-\frac{1}{2}} \left({1+\mathcal{M}^2}\right)^{-2}.\\
\end{split}
\end{equation}
Since the heating timescale is shorter than the cooling timescale everywhere within $r\simeq R_{\rm BHL}$, 
radiative cooling does not play an important role in the accretion flow as long as $\dot{m}_{\rm B}\lesssim7\times10^{-3}$.

The $r_{\rm in}$ dependence of the net accretion rate affects the actual BH feeding rate and radiative output from the nuclear disk at $r\ll r_{\rm in}$.
Numerical simulations of RIAFs find that the positive gradient of the inflow rate (i.e., $s\equiv d\ln \dot{M}/d\ln r>0$) ceases
and the net accretion rate becomes constant within a transition radius of $r_{\rm tr} \approx (30-100)r_{\rm Sch}$ 
\citep[e.g.,][]{Abramowicz2002ApJ...565.1101A,Narayan2012MNRAS.426.3241N, 
Yuan2012ApJ...761..129Y, Sadowski2015MNRAS.447...49S}.
Assuming $s=1/2$, the reduction factor of the net accretion rate is estimated as $\sim (r_{\rm in}/100~r_{\rm Sch})^s \simeq 5$
for a RIAF onto a moving BH with $c_\infty =500~{\rm km~s}^{-1}$ and $\mathcal{M}=2$.
In \app\ref{appendix:spectra}, we discuss how radiation spectra are modified by this effect.

We note that the similarity between MHD and HD simulations seen at larger scales would not hold all the way down to the event horizon scales. In the inner region ($r\ll r_{\rm in}$), since the adiabatic index of gas changes from $5/3$ to $4/3$ because of relativistic effects and cooling processes (synchrotron and/or thermal conduction), magnetic field would be dynamically more important as seen in MHD simulations with general relativistic effects. However, the estimation of the transition scale is beyond our scope in this paper.

%%% Fig.10 %%

%%%%%%%%%%%%%%
%	4. Wandering BH      %
%%%%%%%%%%%%%%
\vspace{5mm}
\section{Radiation spectra of wandering BHs }\label{sec:discussion}

In this section, we calculate the radiation spectral energy distribution (SED) of accretion flows onto a moving BH 
and discuss the detectability of wandering (SM)BHs in different types of galaxies.
The electromagnetic emission and feeding mechanism of a moving BH have both been studied.  Most previous studies have focused on X-ray emission from low-density accretion flows 
\citep[e.g.,][]{Agol2002MNRAS.334..553A, Tsuna2018MNRAS.477..791T, Manshanden2019JCAP...06..026M, Zivancev&Ostriker2020arXiv200406083Z}, by
analogy with low-luminosity AGNs \citep{Ho2008nuclear, Ho2009radiatively}.
However, the radiation spectrum is expected to peak at $\sim 100$ GHz, for which the radio interferometers
such as ALMA and VLA have the highest sensitivity and spatial resolution 
\citep{VLA1980ApJS...44..151T, ALMA2015ApJ...808L...3A}.

Most radiation is generated at the innermost region of the accretion flow near the BH event horizon.
However, because of the limitation of our numerical simulations, we do not address the properties of accreting gas 
within $r_{\rm in}\, (\gg r_{\rm Sch})$, as discussed in \S\ref{sec:r_in}.
Instead, we here calculate the radial distribution of physical quantities adopting a semi-analytical two-temperature disk model, using 
our simulation data as boundary conditions \citep{Manmoto1997ApJ...489..791M,Yuan2000ApJ...537..236Y}.
Using the profiles, we can quantify the radiation spectrum of a RIAF onto a wandering BH embedded in a hot, diffuse plasma.
Although the model includes several free parameters (e.g., the strength of viscosity and the fraction $\delta$ of turbulent dissipation  
that heats the electrons directly) to characterize the disk properties, we choose their parameters so that the relation between the radiative efficiency and BH accretion rate becomes consistent with the efficiency model by \citet{Inayoshi2019transition}.
The model is based on 
the results of MHD simulations that include general relativistic effects and frequency-dependent radiation transport by 
\citet{Ryan2017ApJ...844L..24R} and a semi-analytical model by \citet{Xie2012MNRAS.427.1580X}.
The details of the model are given in Appendix \ref{appendix:spectra}.

In the following, we consider the radiation spectra from wandering BHs that accrete gas at the outskirts of elliptical galaxies, 
the Milky Way, and satellite dwarf galaxies, and we discuss their detectability by ALMA, VLA, and future facilities such as ngVLA.

%%% Fig.11 %%%
\begin{figure}[t]
    \centering
    \includegraphics[width=\linewidth]{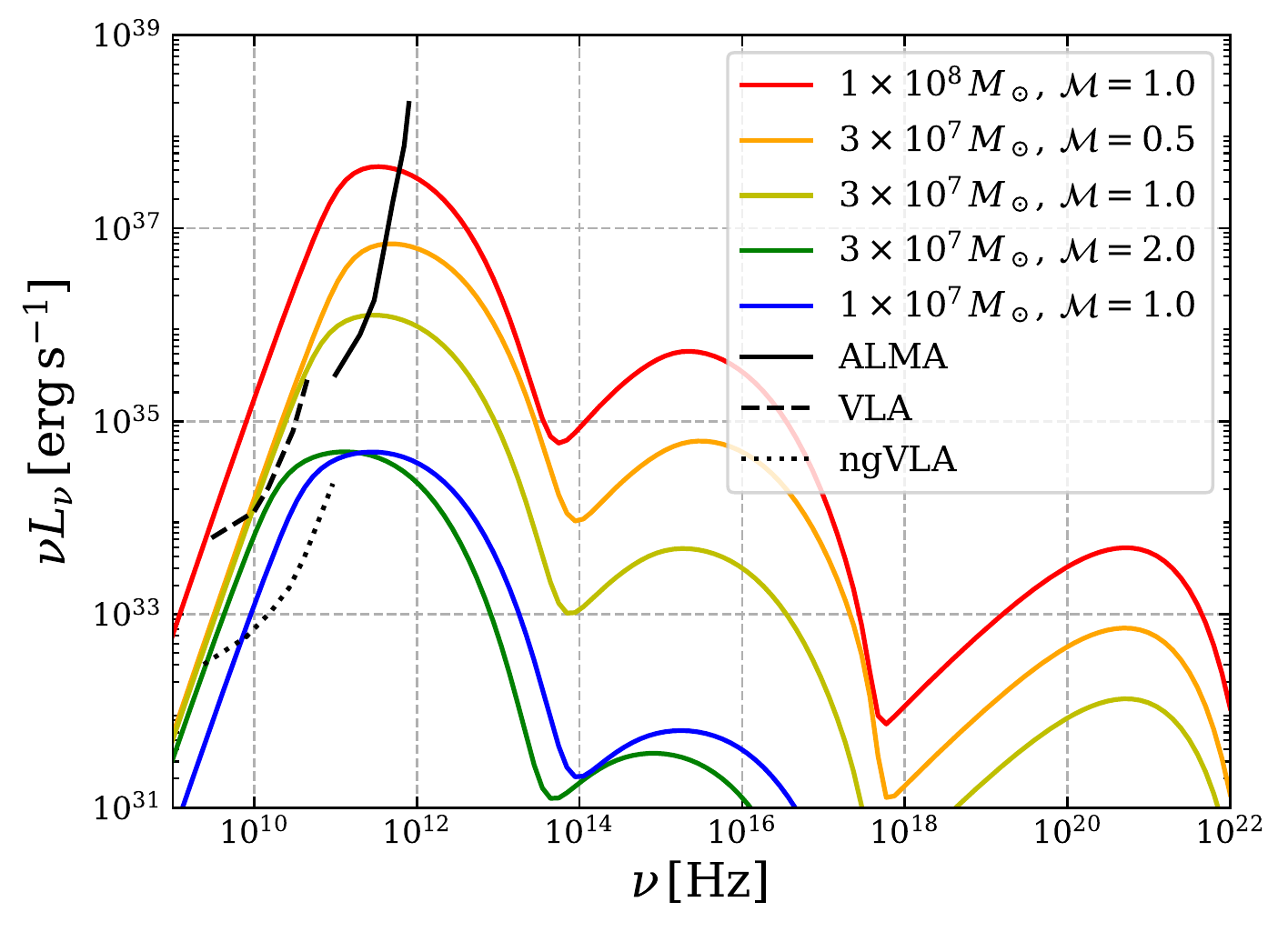}
    \caption{
    Radiation spectra from a wandering BH that accretes hot and diffuse plasma in the outskirts of the massive elliptical galaxy M87.
    Each curve corresponds to a case with different BH mass ($10^7 \leq M_\bullet/\msun \leq 10^8$) and Mach number 
    ($0.5\leq \mathcal{M} \leq 2.0$). 
    The black solid curve shows the ALMA sensitivity to continuum emission, which is manually calculated by the sensitivity calculator provided by the observatory (\url{https://almascience.nrao.edu/proposing/sensitivity-calculator}).
    The sensitivity at each frequency is calculated adopting the rms noise level achieved by single-point 
    1 hour on-source integration (assuming precipitable water vapor of $0.472$ mm).
    The black dashed curve indicates the VLA sensitivity to continuum emission for 1 hour on-source integration, 
    which is manually calculated by the sensitivity calculator provided by the observatory (\url{https://obs.vla.nrao.edu/ect/}).
    The black dotted curve is the ngVLA continuum sensitivity demonstrated by the performance estimates on their website
    (\url{https://ngvla.nrao.edu/page/performance}).
    We set the distance to M87 galaxy ($D=16.68$ Mpc) and adopt the Chandra observational data from 
    \cite{Russell2015MNRAS.451..588R} to model the properties of gas surrounding a wandering BH.
    Wandering BHs with masses of $M_\bullet \gtrsim 3\times 10^7~\rm \msun$, if any,  could be detectable with ALMA and VLA.
    } 
    \label{fig:spectra}
    \vspace{5mm}
\end{figure}

%%% Fig.12 %%%
\begin{figure}[t]
    \centering
    \includegraphics[width=\linewidth]{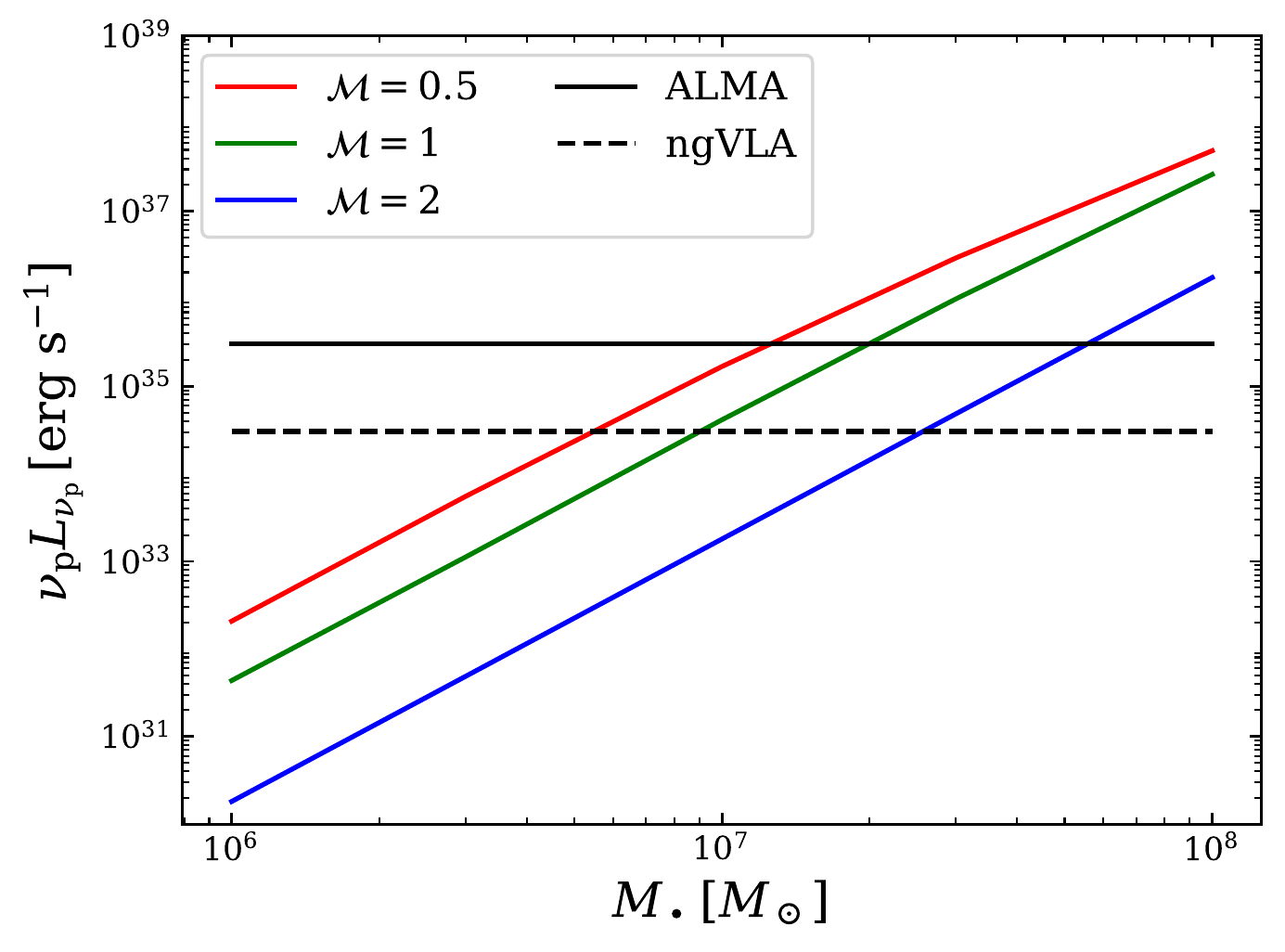}
    \caption{
    Radio luminosity at $100~{\rm GHz}$ for different BH masses and Mach numbers. 
    The horizontal lines are the threshold for detections with the ALMA and ngVLA, assuming the distance to M87.
        } 
    \label{fig:detection}
    \vspace{5mm}
\end{figure}

\subsection{Elliptical galaxies}\label{sec:elliptical}

In the framework of hierarchical structure formation in the $\Lambda$CDM model, lower-mass galaxies form first, and 
they subsequently merge to build larger objects. 
In this paradigm, massive elliptical galaxies in the local universe are expected to experience a large number of galaxy mergers in a Hubble time.
As a natural result of multiple dry mergers at low redshifts (gas-rich mergers at high redshifts), binary SMBHs form at the galactic core,
and some of them merge into a single SMBH through multi-body BH interactions that likely eject the smallest BHs from the core \citep{ Ryu2018MNRAS.473.3410R, Zivancev&Ostriker2020arXiv200406083Z}.
Therefore, some ejected BHs, depending on the kick velocity, are still bound within the galactic halo and orbit at velocities of $v_\infty \sim \sigma_\star$,
where $\sigma_\star$ is the stellar velocity dispersion.
When the orbiting BHs are fed with the diffuse gas of the surrounding host, they emit nonthermal radiation, as discussed below.

%%% Table 2 %%%
\begin{table*}[t]
\caption{\label{tab:detection}
Luminosity of wandering BHs.}
\begin{ruledtabular}
\begin{tabular}{lccccccc}
     Name & $D~{\rm (Mpc)}$ &    $\log{(M_\bullet/\msun)}$ &  $n_{\rm e}~{\rm (cm^{-3})}$ &   $T~{\rm (keV)}$ &  $\log{\dot{m}}$  &  $\log{(L_{\rm bol}/{\rm erg~s^{-1}})}$  & $\log{(F_{\nu_{\rm p}}/{\rm \mu Jy})}$\\
\colrule
 M87 & 16.68 &  9.789$\pm$0.027 &  0.114$\pm$0.016 &  1.650$\pm$0.050 &  $-$5.918$\pm$0.104 &  38.268$\pm$0.220 &   2.685$\pm$0.115 \\
 NGC 507 & 70.80 &  9.210$\pm$0.160 &  0.029$\pm$0.009 &  0.965$\pm$0.015 &  $-$6.742$\pm$0.288 &  36.132$\pm$0.695 &  $-$0.344$\pm$0.618 \\
 NGC 1316 & 20.95 &  8.230$\pm$0.080 &  0.033$\pm$0.007 &  0.620$\pm$0.010 &  $-$7.385$\pm$0.181 &  33.923$\pm$0.437 &  $-$1.551$\pm$0.404 \\
 NGC 4374 & 18.51 &  8.970$\pm$0.050 &  0.022$\pm$0.005 &  0.595$\pm$0.025 &  $-$6.778$\pm$0.157 &  35.703$\pm$0.486 &   0.422$\pm$0.328 \\
 NGC 4472 & 16.72 &  9.400$\pm$0.100 &  0.029$\pm$0.010 &  0.785$\pm$0.005 &  $-$6.418$\pm$0.233 &  36.944$\pm$0.543 &   1.620$\pm$0.393 \\
 NGC 4552 & 15.30 &  8.920$\pm$0.110 &  0.018$\pm$0.004 &  0.455$\pm$0.035 &  $-$6.738$\pm$0.257 &  35.863$\pm$0.592 &   0.621$\pm$0.469 \\
 NGC 4636 & 14.70 &  8.490$\pm$0.080 &  0.028$\pm$0.011 &  0.485$\pm$0.015 &  $-$7.029$\pm$0.244 &  34.879$\pm$0.545 &  $-$0.336$\pm$0.443 \\
 NGC 5044 & 31.20 &  8.710$\pm$0.170 &  0.050$\pm$0.009 &  0.645$\pm$0.015 &  $-$6.748$\pm$0.254 &  35.644$\pm$0.658 &  $-$0.294$\pm$0.534 \\
 NGC 5813 & 32.20 &  8.810$\pm$0.110 &  0.042$\pm$0.009 &  0.585$\pm$0.015 &  $-$6.661$\pm$0.208 &  35.856$\pm$0.563 &  $-$0.085$\pm$0.396 \\
 NGC 5846 & 24.90 &  8.820$\pm$0.110 &  0.042$\pm$0.009 &  0.625$\pm$0.015 &  $-$6.689$\pm$0.210 &  35.859$\pm$0.515 &   0.128$\pm$0.396 \\
\end{tabular}
\tablecomments{Column (1): galaxy name. Column (2): distance. Column (3): mass of the central BH. Columns (4) and (5): electron density and temperature at $\sim2-3~{\rm kpc}$.
Column (6): accretion rate normalized by the Eddington rate ($\dot{m}_{\rm }\equiv{\dot{M}}/{\dot{M}_{\rm Edd}}$) by the scaling relation from the simulation A1, 
assuming that the wandering BH mass is $1\%$ of the central BH mass and $\mathcal{M}=0.5$.
Column (7): bolometric luminosity.
Column (8): radiation flux density at $\nu_{\rm p} = 100~{\rm GHz}$. 
The distance and central BH mass are taken from \citep[][and references therein]{Inayoshi2020ApJ...894..141I}. 
The electron density and temperature are taken from \citet{Russell2015MNRAS.451..588R} for M87 galaxy and from \citet{Russell2013MNRAS.432..530R} for the others.
}
\end{ruledtabular}
    \vspace{5mm}
\end{table*}

As an example of a massive elliptical galaxy, we consider M87.
To model the properties of gas surrounding a wandering BH, we adopt the {\it Chandra} observational 
data from \citet{Russell2015MNRAS.451..588R}:
the electron density $n_e\approx 0.11 ~ {\rm  cm^{-3}}$ ($\rho\approx1.84\times10^{-25}~{\rm g ~ cm^{-3}}$) and 
temperature $T\approx 1.9\times10^7 ~{\rm K}$ ($c_{\rm s}\approx6.5\times10^2~{\rm km~s^{-1}}$) for 
gas at a distance of $r \approx 2-3 ~{\rm kpc}$ from the center.
Since the mass of the central SMBH is as high as $\simeq 6\times 10^9~\msun$ 
\citep{Gebhardt_M87BHmass2011ApJ...729..119G, M87ETH2019ApJ...875L...1E}, the masses of the wandering BHs would be in
the range $M_\bullet \approx 10^{7-8}~\msun$, which corresponds to BH mass ratios of $q\approx 10^{-3}-10^{-2}$. These
mass ratios are reasonable for massive ellipticals that have frequently experienced minor dry mergers (see \figu 1 in \citealt{Ryu2018MNRAS.473.3410R}). 
The orbital velocity of the moving BH is estimated as $v_\infty \sim \sigma_\star \simeq 321~{\rm km}~{\rm s}^{-1}$
(the stellar velocity dispersion is taken from \citealt{Babyk2018ApJ...857...32B}), corresponding to $\mathcal{M}\simeq 0.5$.
Since this estimation is somewhat uncertain and the result is sensitive to the choice of $\mathcal{M}$, as shown below, we treat the Mach number as a free parameter 
in the range of $0.5 \leq \mathcal{M} \leq 2$.
For reference, for a BH with $M_\bullet=3\times10^7~\msun$ moving at a velocity of $\mathcal{M}=1$,
the BH feeding rate is approximated as $\sim 0.2~\dot{M}_{\rm BHL}=2.2\times10^{-7}~\dot{M}_{\rm Edd}$ for the A1 run.

\figu\ref{fig:spectra} presents the radiation spectra with different BH masses of $M_\bullet=10^7$, $3\times10^7$, and $10^8~\msun$ 
and Mach numbers of $\mathcal{M}=0.5$, $1.0$, and $2.0$. 
We also overlay the sensitivity curve of ALMA, assuming a distance of 16.68 Mpc for M87 \citep{Blakeslee2009ApJ...694..556B}.
For all the cases, the radiation spectra have peaks in the millimeter band at $\nu_{\rm p} \simeq 100$ GHz, where the ALMA sensitivity is the highest.
The peak luminosity increases and exceeds the ALMA sensitivity with higher BH masses and lower Mach numbers.

In \figu\ref{fig:detection}, we show the $100$ GHz continuum luminosity as a function of BH mass $M_\bullet$ for different Mach numbers. 
The two horizontal lines correspond to the detection limits for ALMA (solid) and ngVLA (dashed), respectively.
This shows that wandering BHs with $M_\bullet \gtrsim 2\times 10^7~\msun$ and $\mathcal{M}\lesssim 1$ could be detectable with ALMA. 
The detectable BH mass is reduced by a factor of $2-3$ with ngVLA, whose sensitivity is one order of magnitude higher than that of ALMA.
Note that if those BHs are wandering at larger distances of $\sim5-10\,{\rm kpc}$ from the galactic center, where the plasma density is lower,
their luminosities decrease and thus the detection threshold for the BH mass increases by a factor of $\sim 2-4$.

We apply this argument to other nearby massive elliptical galaxies, assuming the existence of wandering BHs 
at their galaxy outskirts.
Taking the observational data from \citet{Russell2013MNRAS.432..530R,Russell2015MNRAS.451..588R}
and \citet{Inayoshi2020ApJ...894..141I}, we estimate the properties of gas surrounding 
those BHs and quantify their predicted bolometric luminosities and 100 GHz flux densities.
The errors of density and temperature are given by the maximum and minimum values 
at distances of $r \simeq 2-3~{\rm kpc}$ from the centers. 
We assume the mass of the wandering BH to be 1\% of the central SMBH mass, and we choose $\mathcal{M}=0.5$ (note that $\sigma_\star/c_{\rm s}\simeq0.5-0.8$ for most cases in our sample).
As shown in \tab\ref{tab:detection}, the bolometric luminosities produced from wandering BHs are on the order of 
$\simeq 10^{35}-10^{-36}~{\rm erg~s^{-1}}$ and the flux densities at $100~{\rm GHz}$ are 
$10^{-1}-10^{3}~{\rm \mu Jy}$. 
We note that the ALMA sensitivity at 100 GHz is $\sim 9~{\rm \mu Jy}$ for 1 hour on-source integration.
Therefore, BHs, if any, wandering at the galactic outskirts could be detectable in M87 and NGC 4472.
With the capability of ALMA, only a few nearby ($\lesssim 20$ Mpc) ellipticals are 
interesting targets for hunting wandering BHs.

Finally, we generalize this argument for early-type, gas-poor galaxies of several morphological types 
and give an estimate of the millimeter luminosity from wandering BHs as a function of the stellar velocity 
dispersion $\sigma_\star$.
To characterize the gas density and temperature of the ambient environment of the wandering BH, 
we approximate the density profile with an isothermal $\beta$-model
\begin{equation}
    \rho=\frac{\rho_0}{(1+r^2/r_{\rm c}^2)^{1.5}}, 
\end{equation}
where $r_{\rm c}$ is the core radius, and the core density and gas temperature are estimated with 
Eqs. (22) and (23) in \citet{Zivancev&Ostriker2020arXiv200406083Z} 
(scaling relations fitted with data from \citealt{Babyk2018ApJ...857...32B}) as
\begin{subequations}
\begin{equation}
 \log \left(\frac{\rho_0}{{\rm g~cm^{-3}}}\right)= 0.6 ~\log \left(\frac{\sigma_\star}{{\rm 100 ~ km~s^{-1}}}\right) - 23.8,
\end{equation}
\begin{equation}
    \log \left(\frac{T}{{\rm keV}} \right)= 2.50~\log \left( \frac{\sigma_\star}{{\rm 100 ~ km~s^{-1}}} \right) - 1.06.
\end{equation}
\end{subequations}
We estimate the mass of the central SMBH using the $M_\bullet-\sigma_\star$ relation 
\citep{Kormendy2013ARA&A..51..511K} and set the mass of the wandering BH to $1\%$ of the nuclear SMBH.
As a reference, the orbital distance of the wandering BH from the galactic center is set to $r=3~r_{\rm c}$, and 
its velocity relative to the surrounding hot gas is set to $\mathcal{M}=0.5$.
We estimate the relation between the luminosity at $\nu_{\rm p}=100~{\rm GHz}$ and the central velocity dispersion as 
\begin{equation}
\log \left( \frac{\nu_{\rm p} L_{\nu_{\rm p}}}{{\rm erg~s^{-1}}} \right) = {7.6~\log\left(\frac{\sigma_\star}{100~{\rm km~s^{-1}}}\right)+32.1}.
\end{equation}
For distances comparable to that of M87, galaxies with $\sigma_\star\gtrsim320 ~{\rm km~s^{-1}}$ yield $\nu_{\rm p}L_{\nu_{\rm p}}\gtrsim10^{36}~{\rm erg~s^{-1}}$, which can be detected by ALMA.

\subsection{Milky Way}

The existence of intermediate-mass BHs (IMBHs; see a recent review by \citealt{Greene_ARAA_2019}) with $M_\bullet \gtrsim 10^4~\msun$ in our Galaxy has been argued based on
observations of high-velocity compact clouds \citep{Oka2017NatAs...1..709O, Tsuboi2017ApJ...850L...5T, Ravi2018MNRAS.478L..72R}
and theoretical/numerical studies \citep{Volonteri2005MNRAS.358..913V, Bellovary2010ApJ...721L.148B, Tremmel2018MNRAS.475.4967T}. 
\citet{Tremmel2018ApJ...857L..22T} predict that Milky Way-size halos would host $\sim 10$ IMBHs within their virial radii, and that they
would be wandering within their host galaxies for several gigayears.

\begin{figure}[t]
    \centering
    \includegraphics[width=\linewidth]{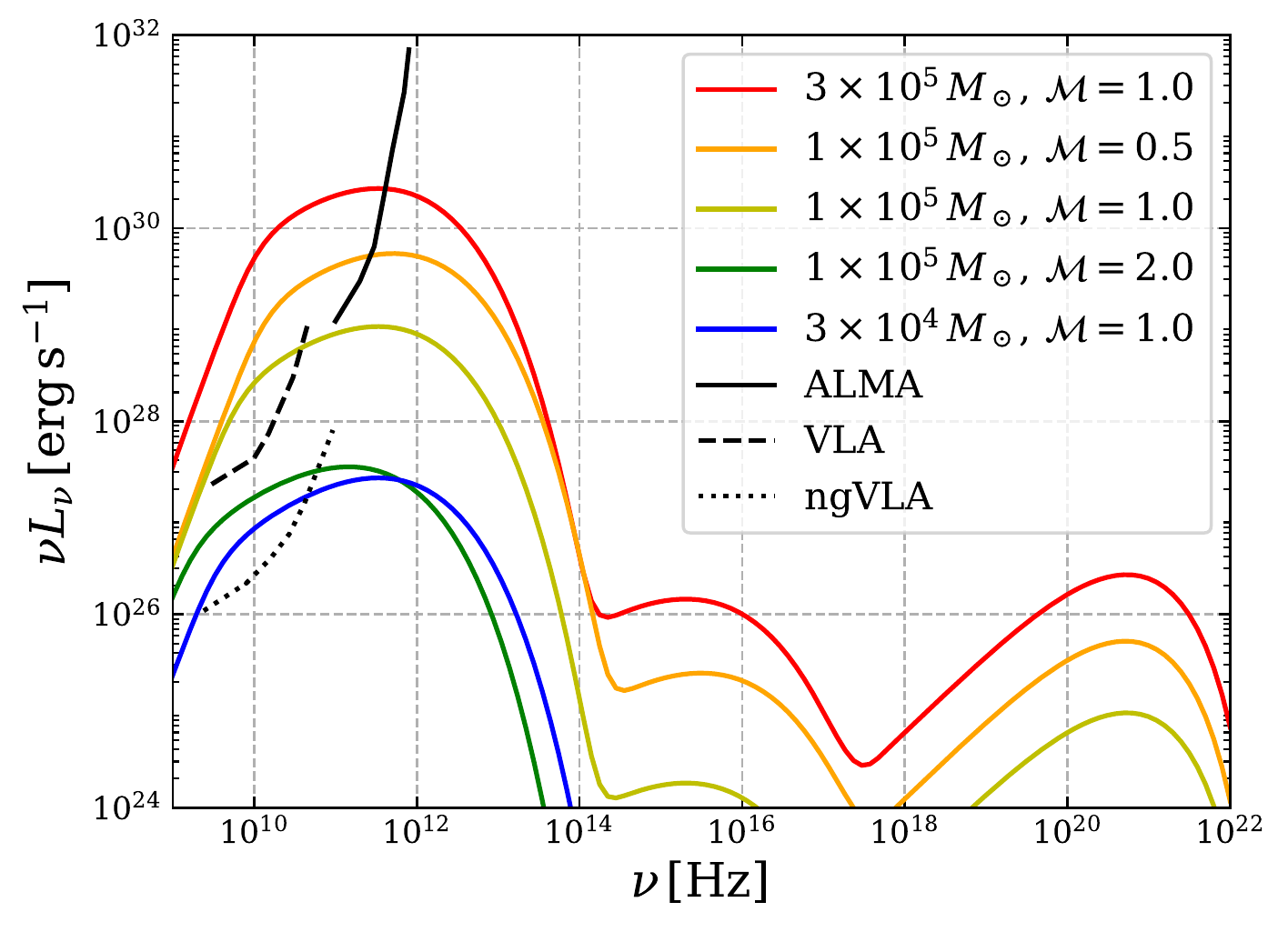}
    \caption{
    Similar to \figu\ref{fig:spectra}, but for accretion flows of wandering BHs in the outskirts of the Milky Way. 
    } 
    \label{fig:spectra_MW}
    \vspace{5mm}
\end{figure}

We apply the same exercise as in \S\ref{sec:elliptical} for wandering BHs with kpc-scale orbits within the Milky Way.  
To model the properties of the hot gas surrounding the Milky Way halo, we adopt the results of the Suzaku X-ray observations 
\citep{Nakashima2018ApJ...862...34N}, which estimate a plasma temperature 
of $T\simeq 3\times 10^6\K$ and an emission measure of $\rm EM \simeq (0.6-16.4)\times 10^{-3}\,{\rm cm}^{-6}~{\rm pc}$.
Based on these results, we adopt $n_{\rm e}= 0.01~\cc$ as the gas density around wandering BHs\footnote{The electron number density is inferred as $n_{\rm e}\approx 4\times 10^{-3}~\cc$ in \cite{Nakashima2018ApJ...862...34N},
assuming spherical and disk-like distributions of gas. 
The value we adopt is higher than the median by a factor of 2.5, but is within the spatial fluctuation of the emission measure.}.
In \figu\ref{fig:spectra_MW}, we show the radiation spectra of wandering BHs with $M_\bullet \approx 3\times 10^4 - 3\times 10^5\,\msun$
located at $\sim 10~{\rm kpc}$ from the Earth.
The spectra in the millimeter band extend to lower frequencies, where (ng)VLA has the highest sensitivity.
We could detect IMBHs down to $M_\bullet \gtrsim 10^5~\msun$ for $\mathcal{M}\lesssim1$. 

There is additional indirect evidence of the existence of hot gas in the Milky Way halo at distances larger than $\sim 50~{\rm kpc}$,
based on observations of the Local Group dwarf galaxies with gas removed by ram pressure stripping \citep{Grcevich2009ApJ...696..385G} and absorption lines of high-velocity clouds associated with the Magellanic Stream that is close to pressure equilibrium with a hot plasma 
\citep{Fox2005ApJ...630..332F}.
Those observations suggest a lower density for the hot gas halo ($n\approx 10^{-4}~{\rm cm^{-3}}$), which, if true, would imply that wandering BHs in the Milky Way halo are too dim to be detected.

%%% Fig.13 %%%

\subsection{Dwarf galaxies}

Observations have identified IMBHs in low-mass dwarf galaxies with confidence down to $M_\bullet \approx 10^5-10^6~\msun$, and more tentatively for 
$M_\bullet \lesssim 10^5~\msun$ (\citealt{Greene_ARAA_2019}; see also \citealt{2017IJMPD..2630021M}, \citealt{Mezcua2020ApJ...898L..30M}).  
Cosmological simulations studying the occupation fraction of IMBHs in dwarf galaxies and find that a significant fraction of them are not centrally located but wander 
within a few kpc from the galaxy centers \citep{2019MNRAS.482.2913B}.
This is expected for dwarf galaxies because of their shallow gravitational potential wells and the longer dynamical friction timescale for the wandering BHs.
Multi-body BH interactions and GW recoils due to BH mergers further contribute
to the off-nuclear population of IMBHs \citep{Lousto2012PhRvD..85h4015L, Bonetti2019MNRAS.486.4044B}.

Recent radio observations of dwarf galaxies with the VLA by \cite{Reines2020ApJ...888...36R} reported a sample of wandering IMBH candidates 
that are significantly offset from the optical centers of the host galaxies. Based on an empirical scaling relation between BH mass and total stellar mass, 
these authors argue that the candidate wandering BHs might have masses in the range $M_\bullet\approx 10^{4}-10^{6}~\msun$.
With radio luminosities of $\sim 10^{20}-10^{22}~{\rm W~Hz^{-1}}$ at $\nu_0 =9~{\rm GHz}$, the sources are radiating at $\gtrsim \nu_0 L_{\nu_0} \simeq 10^{-6}~L_{\rm Edd}$.
Based on the radiative efficiency model for RIAFs, this level of (bolometric) luminosity can be produced only when BHs accrete at relatively high 
accretion rates of $\dot{m}\approx 10^{-4}$.  However, at such a high accretion rate, radio synchrotron photons are heated to X-rays via inverse Compton scattering \citep{Ryan2017ApJ...844L..24R}.  

The brightness of the radio emission could be explained by synchrotron radiation from nonthermal electrons accelerated in a relativistic jet instead of arising from a disk.  Since the majority of the Reines et al. candidate wandering BHs are point-like sources at a resolution of $\sim 0\farcs25$ (which corresponds to a physical scale of $\sim 85$ pc at the median distance of the sources), the jet age can be constrained to $\lesssim 10^3$ yr for an assumed jet propagation speed of $\sim 0.3c$ \citep[e.g.,][]{2006ApJ...648..148N,2008A&A...487..885O}.  The hypothesis of young jets seems consistent with their steep spectral indices ($F_\nu \propto \nu^{-0.79}$), analogous to compact steep-spectrum sources \citep{1998PASP..110..493O}, although the spectral indices were estimated over a narrow frequency range ($9-10.65$ GHz).

\section{Summary} \label{sec:sum}

We perform 3D hydrodynamical simulations to investigate the dynamics of radiatively inefficient gas accretion flows onto massive BHs orbiting around the outskirts of their host galaxies in the presence of a hot and diffuse plasma.  A population of wandering BHs can arise from ejection from the galactic nuclei through multi-body BH interactions and GW recoils associated with galaxy mergers and BH coalescences.  We find that when a wandering BH is fed with hot, diffuse plasma with density fluctuations, the accretion flow 
forms a geometrically thick and hot disk.
Owing to a wide distribution of inflowing angular momentum, 
the mass accretion rate is limited at $\sim 10\%-20\%$ of the canonical Bondi-Hoyle-Littleton rate and decreases as the innermost radius decreases following a power law $r_{\rm in}\,\propto r^{-1/2}$.

Using the simulation results, we further calculate the radiation spectra of the radiatively inefficient accretion flows, which peak in the millimeter band ($\sim 100$ GHz).  We show that the predicted signal may be detectable with ALMA for a hypothetical wandering BH with $M_\bullet \simeq 2\times10^7~\msun$ orbiting a massive ($\sigma_\star\gtrsim 300~{\rm km/s}$) nearby elliptical galaxy such as M87, or $M_\bullet \simeq 10^5~\msun$ moving through the halo of the Milky Way.  The sensitivity will improve with future facilities such as ngVLA.
Our radiation spectral model, combined with numerical simulations, can be applied to provide physical interpretations of candidate off-nuclear BHs detected in nearby dwarf galaxies, which may constrain BH seed formation scenarios.
\\

\acknowledgments
\section*{Acknowledgement}

We greatly thank Feng Yuan, Kengo Tomida, and Kohei Ichikawa for the constructive discussion. 
This work is partially supported by the National Science Foundation of China (11721303, 11991052, 11950410493) and
the National Key R\&D Program of China (2016YFA0400702). 
Numerical computations were carried out with the High-performance Computing Platform of Peking University and 
Cray XC50 at the Center for Computational Astrophysics of the National Astronomical Observatory of Japan.

%%%%%%%%%%%%%%%%%%%%%%%%%%%%%%%%%%%%%%%%%%%%%%%%%%%%%%%%%%%%%%%%%%%%%%%%%%%%%%%%%%%%%%%%%%%%%%%%%%%%%%%%%%%%%%%%

\vspace{5mm}
\appendix

\section{A toy model of the accretion}
\label{appendix:toymodel}

We briefly describe the physical reason why the mass inflow rate within the BHL radius 
scales with $\propto r^{1/2}$, as seen in our simulations.
For simplicity, we consider only the density perturbation along the $y$-axis 
(note that the density gradient to the $z$-axis does not affect the following argument because of its symmetry across 
the $z=0$ plane), and thus the density field at infinity is expressed as 
\begin{equation}
    \rho=\rho_\infty\{1+\sin(ky)\},
\end{equation}
where $k=2\pi/\lambda$ and the amplitude is set to unity.
Let us consider a test particle (or supersonic fluid particle) that has a velocity of 
$\boldsymbol{v} =-v_\infty \boldsymbol{\hat{x}}$ at $x\rightarrow +\infty$
and a distance of $\zeta=\sqrt{y^2+z^2}$ from the $x$-axis,
where ($y,z$) is the position of the particle at infinity.
Defining $\cos \alpha = y/\zeta$, the specific angular momentum of the particle to the $z$-axis is 
given by $j_{\rm z}(\zeta, \alpha)=v_\infty \zeta \cos{\alpha}$.
From a simple analytic calculation, one finds that particles with the same value of $\zeta$ will
collide behind the BH at a distance of $r_0=\rnew^2v_\infty^2/(2GM_\bullet)$.
In the classical picture in \cite{Hoyle1939PCPS...35..405H}, where the density gradient and fluctuation are not considered,
the net angular momentum is set to zero.
On the contrary, with density fluctuations, a non-zero angular momentum is left owing to mass asymmetry:
\begin{equation}
    \bar{j}_z(\rnew)
    =\frac{\int\limits_0^{2\pi}\rho_\infty(1+\sin(k\rnew\cos{\alpha}))v_\infty \rnew\cos{\alpha}{\rm d}\alpha}{\int\limits_0^{2\pi}\rho_\infty(1+\sin(k\rnew\cos{\alpha})){\rm d}\alpha}=v_\infty\rnew J_1(k\rnew),
\label{eq:jzdist}
\end{equation}
where $J_1(x)$ is Bessel function of the first kind, and we obtain 
$\bar{j}_z/j_{\rm Kep}(r_0)=\sqrt{2}~J_1(k\rnew)$.

Let us focus on mass accretion of flows with $\zeta \lesssim R_{\rm BHL}$.
For given $\zeta$, three different types of accretion flows are considered, depending on 
the wavelength of the density fluctuation.
For $k\zeta \ll1$, the system asymptotically approaches the canonical BHL accretion.
For $k\zeta \gg1$, the net angular momentum becomes as small as $\sim O(1/k\zeta)$, but the flow becomes turbulent.
In this case, the accretion rate is close to the BHL rate.
For $k\zeta\lesssim 1$, which corresponds to the case of interest, the net angular momentum 
is approximated as $\bar{j}_z/j_{\rm Kep}\sim k\rnew/\sqrt{2}$, and thus the circularization radius 
is given by $r_{\rm c}=k^2v_\infty^2\rnew^4/(4GM_\bullet)$.
Therefore, the mass inflow rate $\dot{M}_{\rm in}$ through radius $r$ consists of gas flows from 
$\zeta < \zeta_0\equiv \{4GM_\bullet r/(k^2v_\infty^2)\}^{1/4}$ at infinity and is expressed as
\begin{equation}
    \dot{M}_{\rm in}(r_{\rm })=\int\limits_0^{\rnew_0}{\rm d}\rnew\int\limits_0^{2\pi}{\rm d}\alpha ~\zeta \rho v_\infty
    =\pi\rho_\infty v_\infty\rnew_0^2 
    =\frac{1}{\sqrt{2}}~\rho_\infty \lambda r v_{\rm ff}\propto r^{1/2},
\end{equation}
where $v_{\rm ff}=\sqrt{2GM_\bullet /r}$.
Equating this to $4\pi \rho r^2 v_{\rm ff}$, we obtain 
\begin{equation}
\rho = \frac{\rho_\infty}{2\sqrt{2}~kR_{\rm BHL}}\left(\frac{r}{R_{\rm BHL}}\right)^{-1}.
\end{equation}
Note that $\zeta \lesssim R_{\rm BHL}$ and $k\zeta \lesssim 1$. 
Therefore, the density distribution is approximated as $\rho \sim \rho_\infty (r/R_{\rm BHL})^{-1}$.

In contrast, when the wavelength is smaller than the BH influence radius, gas with lower,
even negative ($\bar{j}_z<0$), angular momentum supplies a substantial fraction of mass, as shown in the case of the B1 run.

\section{The model of disk emission}
\label{appendix:spectra}

To calculate radiation spectra from accretion flows around a BH, we solve the following equations 
to construct the dynamical structure of two-temperature RIAFs
\citep{Nakamura1997PASJ...49..503N,Manmoto1997ApJ...489..791M,Xie2012MNRAS.427.1580X}:
\begin{equation}
    \dot{M}=-4\pi r \rho H v_r, 
\end{equation}
\begin{equation}
    v_r\frac{{\rm d}v_r}{{\rm d}r}=\Omega^2r-\Omega^2_{\rm K}r - \frac{1}{\rho}\frac{{\rm d}p}{{\rm d}r},
\end{equation}
\begin{equation}
    v_r(\Omega r^2 - j) = -\alpha r \frac{p}{\rho},
\end{equation}
\begin{equation}
    \rho v_r \Big{(} \frac{{\rm d}\epsilon_{\rm i}}{{\rm d}r} - \frac{p_{\rm i}}{\rho^2} \frac{{\rm d} \rho}{{\rm d}r}  \Big{)} = (1-\delta)q_{\rm vis} - q_{\rm ie},
\end{equation}
\begin{equation}
    \rho v_r \Big{(} \frac{{\rm d}\epsilon_{\rm e}}{{\rm d}r} - \frac{p_{\rm e}}{\rho^2} \frac{{\rm d} \rho}{{\rm d}r}  \Big{)} = \delta q_{\rm vis} + q_{\rm ie} - q_{\rm rad},
\end{equation}
where $H=c_{\rm s}/\Omega_{\rm K}$ is the disk scale height, 
$\Omega$ is the angular velocity, $\Omega_{\rm K}$ is the Keplerian angular velocity on the equatorial plane, $j$ is the eigenvalue of the equations, $\alpha$ is the viscous parameter, $\epsilon$ is the specific internal energy, $q_{\rm vis}\equiv-\alpha p r \frac{{\rm d}\Omega}{{\rm d}r} $ is the heating rate due to viscous dissipation, $q_{\rm ie}$ is the energy transfer rate by Coulomb collisions between ions and electrons, and $q_{\rm rad}$ is the radiative cooling rate that includes 
synchrotron, bremsstrahlung, and inverse Compton scattering. 
We use the pseudo-gravitational potential $\phi=-GM/(r-r_{\rm Sch})$ \citep{Paczynsky1980A&A....88...23P}. 
The pressure is given by the sum of gas pressure and magnetic pressure ($p=p_{\rm gas}+p_{\rm mag}$), 
where the magnetic pressure is set by assuming the global plasma-$\beta$ value of $\beta=p_{\rm gas}/p$.
The subscripts ``i'' and ``e'' denote physical quantities of ions and electrons, respectively. 
Following the numerical procedure in \citet{Nakamura1997PASJ...49..503N}, the equations are first reduced to 
a set of differential equations of $v,T_{\rm i}$ and $T_{\rm e}$. 
Given the values of the global parameters $M,\,\dot{M},\,\alpha,\,\beta,\,\delta,\,s$ and outer boundary values of $v_r,T_{\rm i},T_{\rm e}$, we numerically solve those equations and find the physical global solution by adjusting the eigenvalue $j$.

\begin{figure}[t]
    \centering
    \includegraphics[width=0.6\linewidth]{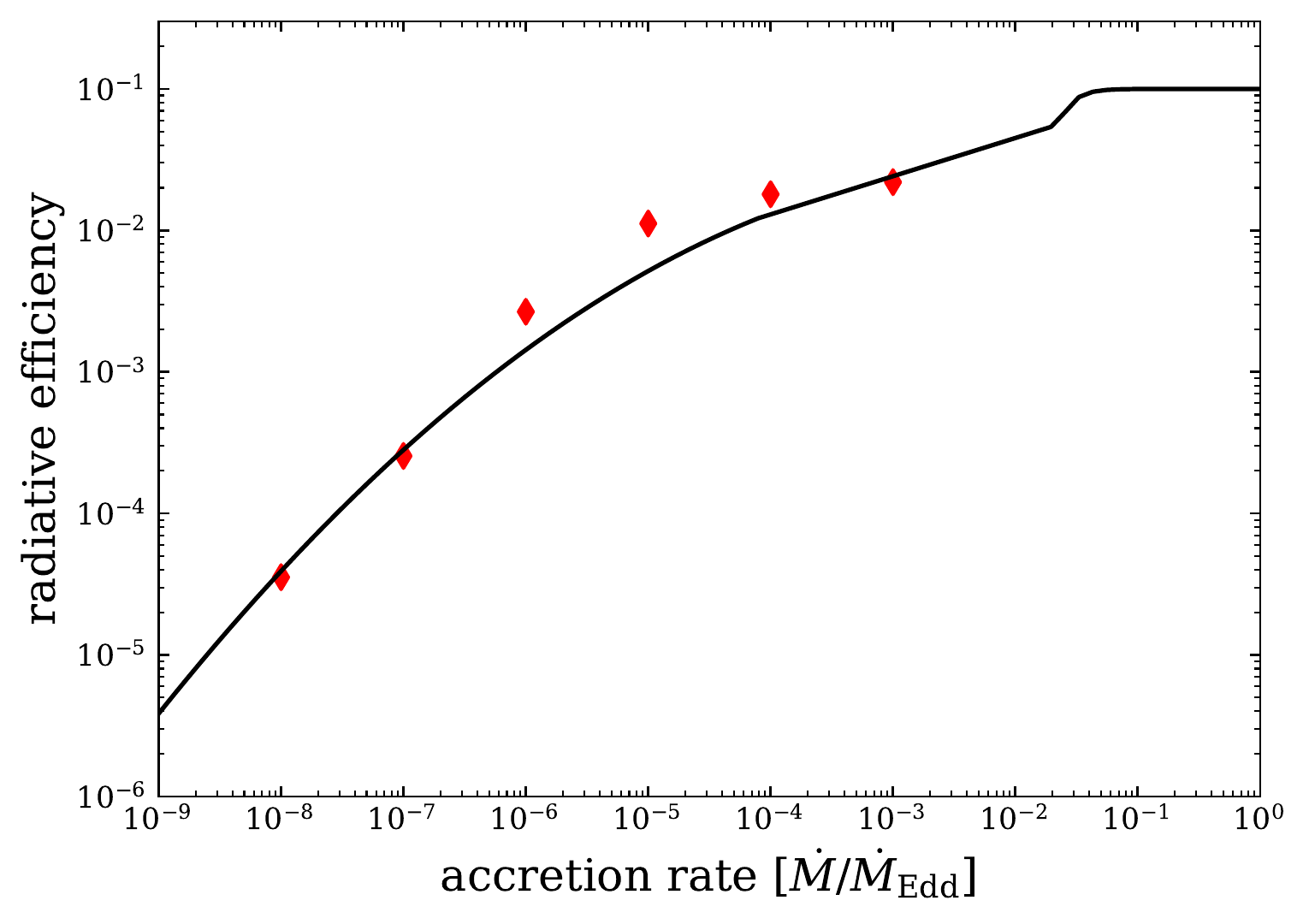}
    \caption{
    Radiative efficiency for RIAFs as a function of the BH accretion rate in units of $\dot{M}_{\rm Edd}$ 
    (solid curve; see more details in \citealt{Inayoshi2019transition}).
    The red symbols present the results obtained from our radiation spectral model, where 
     $\alpha=0.1,\beta=0.8$, and $\delta=0.9$.
    } 
    \label{fig:emdot}
\end{figure}

\begin{figure}[t]
    \centering
    \includegraphics[width=0.6\linewidth]{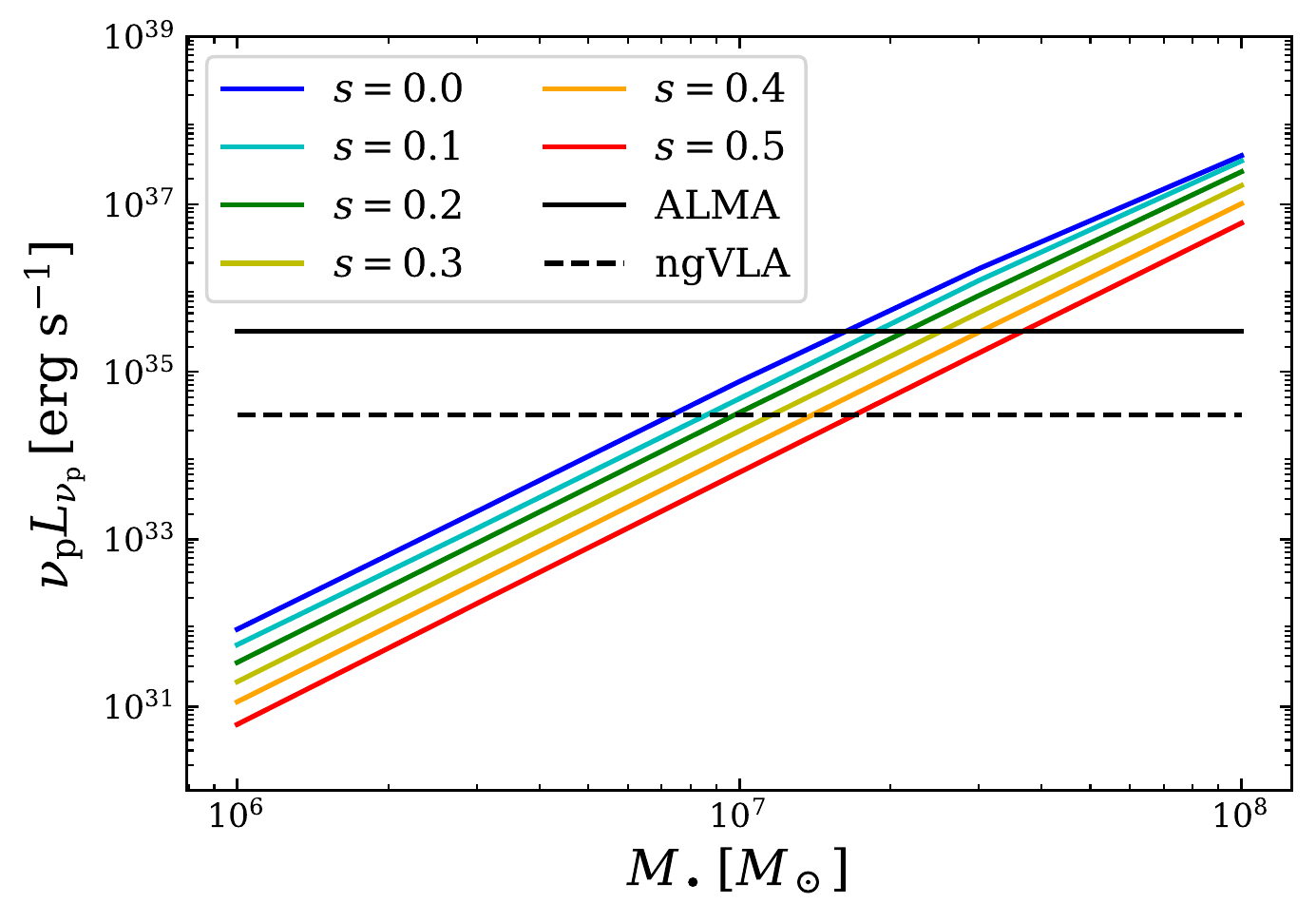}
    \caption{
    Luminosity at $100~{\rm GHz}$ for various values of $s(\equiv d\ln \dot{M}/d\ln r)$ as a function of BH mass. 
    The model parameters are the same as in the case with $\mathcal{M}=1$ shown in \figu\ref{fig:detection}. 
    The $100$-GHz luminosity is reduced by one order of magnitude for $s\simeq 0.5$.
    } 
    \label{fig:detection_s}
\end{figure}

With the flow solution, we calculate the radiative flux from the flow, taking into account 
synchrotron, bremsstrahlung, and inverse Compton scattering for the calculation of spectrum 
\citep[see more details in][]{Manmoto1997ApJ...489..791M}. 
The unscattered spectrum at a distance of $r$ is given by
\begin{equation}
    F_\nu(r)=\frac{2\pi}{\sqrt{3}}B_\nu\left[1-\exp\left( \sqrt{3\pi}\kappa_\nu H\right)\right],
\end{equation}
where $\kappa_\nu=\chi_\nu/(4\pi B_\nu)$ and 
$\chi_\nu=\chi_{\nu,\,\rm brems}+\chi_{\nu,\,\rm synch}$ is the emissivity. 
Using the formula given by \citet{Coppi1990MNRAS.245..453C}, we calculate the Compton-scattered spectrum.
In integrating the flux over the disk, we consider the Doppler effect of emerging photons (observed from infinity)
due to the BH gravity, as $I_{\rm obs}=(\nu_{\rm obs}/\nu)^3 I_\nu$, 
where $\nu_{\rm obs}/\nu=\sqrt{1-(r_{\rm Sch}/r)}$.

In this model, there are three free parameters ($\alpha,\beta$ and $\delta$) to characterize the disk properties.
We calibrate these parameters so that the radiative efficiency for a RIAF modeled by \citet{Inayoshi2019transition},
based on GRRMHD simulations \citep{Ryan2017ApJ...844L..24R} and semi-analytical calculations
\citep{Xie2012MNRAS.427.1580X}, is reproduced as shown in \figu\ref{fig:emdot}.
In our calculation in \S\ref{sec:discussion}, these parameters are set to $\alpha=0.1,\beta=0.8$, and $\delta=0.9$. 
We also use the values at the inner boundary of our simulations as the outer boundary conditions of the calculations.

Finally, we generalize the radiation spectral model, considering the radial-dependent mass accretion rate,
\begin{equation}
    \dot{M}(r)=\dot{M}_0\left(\frac{r}{r_{\rm in}}\right)^s,
\end{equation}
where $r_{\rm in}$ is the location of the innermost cells in our simulation domain
and $\dot{M}_0$ is the mass accretion rate at $r=r_{\rm in}$.
In \S\ref{sec:discussion}, we assume a constant accretion rate (i.e., $s=0$). 
As shown in Figure \ref{fig:massflux_rin}, however, the mass inflow rate decreases toward the center, 
indicating $s\simeq 0.5$.
Numerical simulations of RIAFs also show that a positive gradient of the inflow rate ceases
and the net accretion rate becomes constant within a transition radius 
$r_{\rm tr} \approx  (30-100)~r_{\rm Sch}$ \citep[e.g.,][]{Abramowicz2002ApJ...565.1101A,
Narayan2012MNRAS.426.3241N, Yuan2012ApJ...761..129Y,Sadowski2015MNRAS.447...49S}.
We study the dependence of radiation spectra on the value of $s$. 
In \figu\ref{fig:detection_s}, the luminosities at $100$ GHz for various values of $0\leq s\leq 0.5$ are shown,
where $r_{\rm in}=0.08~R_{\rm BHL}$, $r_{\rm tr}=100~r_{\rm Sch}$, and $\mathcal{M}=1$ are adopted. 
As a result of the reduction of accreted mass, the radio luminosity decreases as 
$\propto (r_{\rm tr}/r_{\rm in})^{2s} \sim 10^{-2s}$, which increases the detectable mass of 
wandering BHs by a factor of $\sim 3$ for $s=0.5$

\bibliography{ref} 

\begin{thebibliography}{}
\expandafter\ifx\csname natexlab\endcsname\relax\def\natexlab#1{#1}\fi
\providecommand{\url}[1]{\href{#1}{#1}}

\bibitem[{{Abramowicz} {et~al.}(2002){Abramowicz}, {Igumenshchev}, {Quataert},
  \& {Narayan}}]{Abramowicz2002ApJ...565.1101A}
{Abramowicz}, M.~A., {Igumenshchev}, I.~V., {Quataert}, E., \& {Narayan}, R.
  2002, \apj, 565, 1101

\bibitem[{{Agol} \& {Kamionkowski}(2002)}]{Agol2002MNRAS.334..553A}
{Agol}, E., \& {Kamionkowski}, M. 2002, \mnras, 334, 553

\bibitem[{{ALMA Partnership} {et~al.}(2015){ALMA Partnership}, {Brogan},
  {P{\'e}rez}, {Hunter}, {Dent}, {Hales}, {Hills}, {Corder}, {Fomalont},
  {Vlahakis}, {Asaki}, {Barkats}, {Hirota}, {Hodge}, {Impellizzeri}, {Kneissl},
  {Liuzzo}, {Lucas}, {Marcelino}, {Matsushita}, {Nakanishi}, {Phillips},
  {Richards}, {Toledo}, {Aladro}, {Broguiere}, {Cortes}, {Cortes}, {Espada},
  {Galarza}, {Garcia-Appadoo}, {Guzman-Ramirez}, {Humphreys}, {Jung}, {Kameno},
  {Laing}, {Leon}, {Marconi}, {Mignano}, {Nikolic}, {Nyman}, {Radiszcz},
  {Remijan}, {Rod{\'o}n}, {Sawada}, {Takahashi}, {Tilanus}, {Vila Vilaro},
  {Watson}, {Wiklind}, {Akiyama}, {Chapillon}, {de Gregorio-Monsalvo}, {Di
  Francesco}, {Gueth}, {Kawamura}, {Lee}, {Nguyen Luong}, {Mangum}, {Pietu},
  {Sanhueza}, {Saigo}, {Takakuwa}, {Ubach}, {van Kempen}, {Wootten},
  {Castro-Carrizo}, {Francke}, {Gallardo}, {Garcia}, {Gonzalez}, {Hill},
  {Kaminski}, {Kurono}, {Liu}, {Lopez}, {Morales}, {Plarre}, {Schieven},
  {Testi}, {Videla}, {Villard}, {Andreani}, {Hibbard}, \&
  {Tatematsu}}]{ALMA2015ApJ...808L...3A}
{ALMA Partnership}, {Brogan}, C.~L., {P{\'e}rez}, L.~M., {et~al.} 2015, \apjl,
  808, L3

\bibitem[{{Babyk} {et~al.}(2018){Babyk}, {McNamara}, {Nulsen}, {Hogan},
  {Vantyghem}, {Russell}, {Pulido}, \& {Edge}}]{Babyk2018ApJ...857...32B}
{Babyk}, I.~V., {McNamara}, B.~R., {Nulsen}, P.~E.~J., {et~al.} 2018, \apj,
  857, 32

\bibitem[{{Begelman} {et~al.}(1980){Begelman}, {Blandford}, \&
  {Rees}}]{Begelman1980Natur.287..307B}
{Begelman}, M.~C., {Blandford}, R.~D., \& {Rees}, M.~J. 1980, \nat, 287, 307

\bibitem[{{Bekenstein}(1973)}]{Bekenstein1973ApJ...183..657B}
{Bekenstein}, J.~D. 1973, \apj, 183, 657

\bibitem[{{Bellovary} {et~al.}(2019){Bellovary}, {Cleary}, {Munshi}, {Tremmel},
  {Christensen}, {Brooks}, \& {Quinn}}]{2019MNRAS.482.2913B}
{Bellovary}, J.~M., {Cleary}, C.~E., {Munshi}, F., {et~al.} 2019, \mnras, 482,
  2913

\bibitem[{{Bellovary} {et~al.}(2010){Bellovary}, {Governato}, {Quinn},
  {Wadsley}, {Shen}, \& {Volonteri}}]{Bellovary2010ApJ...721L.148B}
{Bellovary}, J.~M., {Governato}, F., {Quinn}, T.~R., {et~al.} 2010, \apjl, 721,
  L148

\bibitem[{{Blakeslee} {et~al.}(2009){Blakeslee}, {Jord{\'a}n}, {Mei},
  {C{\^o}t{\'e}}, {Ferrarese}, {Infante}, {Peng}, {Tonry}, \&
  {West}}]{Blakeslee2009ApJ...694..556B}
{Blakeslee}, J.~P., {Jord{\'a}n}, A., {Mei}, S., {et~al.} 2009, \apj, 694, 556

\bibitem[{{Blandford} \&
  {Begelman}(1999)}]{Blandford&Begelman1999MNRAS.303L...1B}
{Blandford}, R.~D., \& {Begelman}, M.~C. 1999, \mnras, 303, L1

\bibitem[{{Blandford} \&
  {Begelman}(2004)}]{Blandford&Begelman2004MNRAS.349...68B}
---. 2004, \mnras, 349, 68

\bibitem[{{Bondi}(1952)}]{Bondi1952spherically}
{Bondi}, H. 1952, \mnras, 112, 195

\bibitem[{{Bonetti} {et~al.}(2018{\natexlab{a}}){Bonetti}, {Haardt}, {Sesana},
  \& {Barausse}}]{Bonetti2018MNRAS.477.3910B}
{Bonetti}, M., {Haardt}, F., {Sesana}, A., \& {Barausse}, E.
  2018{\natexlab{a}}, \mnras, 477, 3910

\bibitem[{{Bonetti} {et~al.}(2018{\natexlab{b}}){Bonetti}, {Sesana},
  {Barausse}, \& {Haardt}}]{Bonetti2018MNRAS.477.2599B}
{Bonetti}, M., {Sesana}, A., {Barausse}, E., \& {Haardt}, F.
  2018{\natexlab{b}}, \mnras, 477, 2599

\bibitem[{{Bonetti} {et~al.}(2019){Bonetti}, {Sesana}, {Haardt}, {Barausse}, \&
  {Colpi}}]{Bonetti2019MNRAS.486.4044B}
{Bonetti}, M., {Sesana}, A., {Haardt}, F., {Barausse}, E., \& {Colpi}, M. 2019,
  \mnras, 486, 4044

\bibitem[{{Campanelli}(2005)}]{Campanelli2005CQGra..22S.387C}
{Campanelli}, M. 2005, Classical and Quantum Gravity, 22, S387

\bibitem[{{Campanelli} {et~al.}(2007{\natexlab{a}}){Campanelli}, {Lousto},
  {Zlochower}, \& {Merritt}}]{Campanelli2007ApJ...659L...5C}
{Campanelli}, M., {Lousto}, C., {Zlochower}, Y., \& {Merritt}, D.
  2007{\natexlab{a}}, \apjl, 659, L5

\bibitem[{{Campanelli} {et~al.}(2007{\natexlab{b}}){Campanelli}, {Lousto},
  {Zlochower}, \& {Merritt}}]{Campanelli2007PhRvL..98w1102C}
{Campanelli}, M., {Lousto}, C.~O., {Zlochower}, Y., \& {Merritt}, D.
  2007{\natexlab{b}}, \prl, 98, 231102

\bibitem[{{Coppi} \& {Blandford}(1990)}]{Coppi1990MNRAS.245..453C}
{Coppi}, P.~S., \& {Blandford}, R.~D. 1990, \mnras, 245, 453

\bibitem[{{Edgar}(2004)}]{Edgar2004NewAR..48..843E}
{Edgar}, R. 2004, \nar, 48, 843

\bibitem[{{Event Horizon Telescope Collaboration} {et~al.}(2019){Event Horizon
  Telescope Collaboration}, {Akiyama}, {Alberdi}, {Alef}, {Asada}, {Azulay},
  {Baczko}, {Ball}, {Balokovi{\'c}}, \& {Barrett}}]{M87ETH2019ApJ...875L...1E}
{Event Horizon Telescope Collaboration}, {Akiyama}, K., {Alberdi}, A., {et~al.}
  2019, \apjl, 875, L1

\bibitem[{{Falcke} \& {Markoff}(2000)}]{Falcke2000A&A...362..113F}
{Falcke}, H., \& {Markoff}, S. 2000, \aap, 362, 113

\bibitem[{{Fox} {et~al.}(2005){Fox}, {Wakker}, {Savage}, {Tripp}, {Sembach}, \&
  {Bland-Hawthorn}}]{Fox2005ApJ...630..332F}
{Fox}, A.~J., {Wakker}, B.~P., {Savage}, B.~D., {et~al.} 2005, \apj, 630, 332

\bibitem[{{Fragione} \& {Silk}(2020)}]{Fragione2020arXiv200601867F}
{Fragione}, G., \& {Silk}, J. 2020, arXiv e-prints, arXiv:2006.01867

\bibitem[{{Fujita}(2008)}]{Fujita2008ApJ...685L..59F}
{Fujita}, Y. 2008, \apjl, 685, L59

\bibitem[{{Fujita}(2009)}]{Fujita2009ApJ...691.1050F}
---. 2009, \apj, 691, 1050

\bibitem[{{Gebhardt} {et~al.}(2011){Gebhardt}, {Adams}, {Richstone}, {Lauer},
  {Faber}, {G{\"u}ltekin}, {Murphy}, \&
  {Tremaine}}]{Gebhardt_M87BHmass2011ApJ...729..119G}
{Gebhardt}, K., {Adams}, J., {Richstone}, D., {et~al.} 2011, \apj, 729, 119

\bibitem[{{Grcevich} \& {Putman}(2009)}]{Grcevich2009ApJ...696..385G}
{Grcevich}, J., \& {Putman}, M.~E. 2009, \apj, 696, 385

\bibitem[{{Greene} {et~al.}(2020){Greene}, {Strader}, \&
  {Ho}}]{Greene_ARAA_2019}
{Greene}, J.~E., {Strader}, J., \& {Ho}, L.~C. 2020, ARA\&A in press,
  arXiv:1911.09678

\bibitem[{{Ho}(1999)}]{Ho1999ApJ...516..672H}
{Ho}, L.~C. 1999, \apj, 516, 672

\bibitem[{{Ho}(2002)}]{Ho2002ApJ...564..120H}
---. 2002, \apj, 564, 120

\bibitem[{{Ho}(2008)}]{Ho2008nuclear}
---. 2008, \araa, 46, 475

\bibitem[{{Ho}(2009)}]{Ho2009radiatively}
---. 2009, \apj, 699, 626

\bibitem[{{Hoyle} \& {Lyttleton}(1939)}]{Hoyle1939PCPS...35..405H}
{Hoyle}, F., \& {Lyttleton}, R.~A. 1939, Proceedings of the Cambridge
  Philosophical Society, 35, 405

\bibitem[{{Ichimaru}(1977)}]{Ichimaru1977ApJ...214..840I}
{Ichimaru}, S. 1977, \apj, 214, 840

\bibitem[{{Igumenshchev} \&
  {Abramowicz}(2000)}]{Igumenshchev2000ApJS..130..463I}
{Igumenshchev}, I.~V., \& {Abramowicz}, M.~A. 2000, \apjs, 130, 463

\bibitem[{{Igumenshchev} {et~al.}(2000){Igumenshchev}, {Abramowicz}, \&
  {Narayan}}]{Igumenshchev2000ApJ...537L..27I}
{Igumenshchev}, I.~V., {Abramowicz}, M.~A., \& {Narayan}, R. 2000, \apjl, 537,
  L27

\bibitem[{{Igumenshchev} {et~al.}(2003){Igumenshchev}, {Narayan}, \&
  {Abramowicz}}]{Igumenshchev2003ApJ...592.1042I}
{Igumenshchev}, I.~V., {Narayan}, R., \& {Abramowicz}, M.~A. 2003, \apj, 592,
  1042

\bibitem[{{Inayoshi} {et~al.}(2018{\natexlab{a}}){Inayoshi}, {Ichikawa}, \&
  {Haiman}}]{Inayoshi2018ApJ...863L..36I}
{Inayoshi}, K., {Ichikawa}, K., \& {Haiman}, Z. 2018{\natexlab{a}}, \apjl, 863,
  L36

\bibitem[{{Inayoshi} {et~al.}(2020){Inayoshi}, {Ichikawa}, \&
  {Ho}}]{Inayoshi2020ApJ...894..141I}
{Inayoshi}, K., {Ichikawa}, K., \& {Ho}, L.~C. 2020, \apj, 894, 141

\bibitem[{{Inayoshi} {et~al.}(2019){Inayoshi}, {Ichikawa}, {Ostriker}, \&
  {Kuiper}}]{Inayoshi2019transition}
{Inayoshi}, K., {Ichikawa}, K., {Ostriker}, J.~P., \& {Kuiper}, R. 2019,
  \mnras, 486, 5377

\bibitem[{{Inayoshi} {et~al.}(2018{\natexlab{b}}){Inayoshi}, {Ostriker},
  {Haiman}, \& {Kuiper}}]{Inayoshi2018low}
{Inayoshi}, K., {Ostriker}, J.~P., {Haiman}, Z., \& {Kuiper}, R.
  2018{\natexlab{b}}, \mnras, 476, 1412

\bibitem[{{Kelley} {et~al.}(2017){Kelley}, {Blecha}, \&
  {Hernquist}}]{Kelley2017MNRAS.464.3131K}
{Kelley}, L.~Z., {Blecha}, L., \& {Hernquist}, L. 2017, \mnras, 464, 3131

\bibitem[{{Khan} {et~al.}(2016){Khan}, {Husa}, {Hannam}, {Ohme}, {P{\"u}rrer},
  {Forteza}, \& {Boh{\'e}}}]{Khan2016PhRvD..93d4007K}
{Khan}, S., {Husa}, S., {Hannam}, M., {et~al.} 2016, \prd, 93, 044007

\bibitem[{{Kormendy} \& {Ho}(2013)}]{Kormendy2013ARA&A..51..511K}
{Kormendy}, J., \& {Ho}, L.~C. 2013, \araa, 51, 511

\bibitem[{{Lousto} {et~al.}(2012){Lousto}, {Zlochower}, {Dotti}, \&
  {Volonteri}}]{Lousto2012PhRvD..85h4015L}
{Lousto}, C.~O., {Zlochower}, Y., {Dotti}, M., \& {Volonteri}, M. 2012, \prd,
  85, 084015

\bibitem[{{Luo} {et~al.}(2016){Luo}, {Chen}, {Duan}, {Gong}, {Hu}, {Ji}, {Liu},
  {Mei}, {Milyukov}, {Sazhin}, {Shao}, {Toth}, {Tu}, {Wang}, {Wang}, {Yeh},
  {Zhan}, {Zhang}, {Zharov}, \& {Zhou}}]{Luo2016CQGra..33c5010L}
{Luo}, J., {Chen}, L.-S., {Duan}, H.-Z., {et~al.} 2016, Classical and Quantum
  Gravity, 33, 035010

\bibitem[{{Mahadevan}(1997)}]{Mahadevan1997ApJ...477..585M}
{Mahadevan}, R. 1997, \apj, 477, 585

\bibitem[{{Manmoto} {et~al.}(1997){Manmoto}, {Mineshige}, \&
  {Kusunose}}]{Manmoto1997ApJ...489..791M}
{Manmoto}, T., {Mineshige}, S., \& {Kusunose}, M. 1997, \apj, 489, 791

\bibitem[{{Manshanden} {et~al.}(2019){Manshanden}, {Gaggero}, {Bertone},
  {Connors}, \& {Ricotti}}]{Manshanden2019JCAP...06..026M}
{Manshanden}, J., {Gaggero}, D., {Bertone}, G., {Connors}, R. M.~T., \&
  {Ricotti}, M. 2019, \jcap, 2019, 026

\bibitem[{{Merritt}(2013)}]{Merritt2013CQGra..30x4005M}
{Merritt}, D. 2013, Classical and Quantum Gravity, 30, 244005

\bibitem[{{Mezcua}(2017)}]{2017IJMPD..2630021M}
{Mezcua}, M. 2017, International Journal of Modern Physics D, 26, 1730021

\bibitem[{{Mezcua} \& {Dom{\'\i}nguez
  S{\'a}nchez}(2020)}]{Mezcua2020ApJ...898L..30M}
{Mezcua}, M., \& {Dom{\'\i}nguez S{\'a}nchez}, H. 2020, \apjl, 898, L30

\bibitem[{{Mignone} {et~al.}(2007){Mignone}, {Bodo}, {Massaglia}, {Matsakos},
  {Tesileanu}, {Zanni}, \& {Ferrari}}]{Mignone2007PLUTO}
{Mignone}, A., {Bodo}, G., {Massaglia}, S., {et~al.} 2007, \apjs, 170, 228

\bibitem[{{Mo{\'s}cibrodzka} {et~al.}(2011){Mo{\'s}cibrodzka}, {Gammie},
  {Dolence}, \& {Shiokawa}}]{Moscibrodzka2011ApJ...735....9M}
{Mo{\'s}cibrodzka}, M., {Gammie}, C.~F., {Dolence}, J.~C., \& {Shiokawa}, H.
  2011, \apj, 735, 9

\bibitem[{{Nagai} {et~al.}(2006){Nagai}, {Inoue}, {Asada}, {Kameno}, \&
  {Doi}}]{2006ApJ...648..148N}
{Nagai}, H., {Inoue}, M., {Asada}, K., {Kameno}, S., \& {Doi}, A. 2006, \apj,
  648, 148

\bibitem[{{Nakamura} {et~al.}(1997){Nakamura}, {Kusunose}, {Matsumoto}, \&
  {Kato}}]{Nakamura1997PASJ...49..503N}
{Nakamura}, K.~E., {Kusunose}, M., {Matsumoto}, R., \& {Kato}, S. 1997, \pasj,
  49, 503

\bibitem[{{Nakashima} {et~al.}(2018){Nakashima}, {Inoue}, {Yamasaki}, {Sofue},
  {Kataoka}, \& {Sakai}}]{Nakashima2018ApJ...862...34N}
{Nakashima}, S., {Inoue}, Y., {Yamasaki}, N., {et~al.} 2018, \apj, 862, 34

\bibitem[{{Narayan} {et~al.}(2000){Narayan}, {Igumenshchev}, \&
  {Abramowicz}}]{Narayan2000CDAF_ApJ...539..798N}
{Narayan}, R., {Igumenshchev}, I.~V., \& {Abramowicz}, M.~A. 2000, \apj, 539,
  798

\bibitem[{{Narayan} {et~al.}(1998){Narayan}, {Mahadevan}, {Grindlay}, {Popham},
  \& {Gammie}}]{Narayan1998ApJ...492..554N}
{Narayan}, R., {Mahadevan}, R., {Grindlay}, J.~E., {Popham}, R.~G., \&
  {Gammie}, C. 1998, \apj, 492, 554

\bibitem[{{Narayan} {et~al.}(2012){Narayan}, {S{\"A} dowski}, {Penna}, \&
  {Kulkarni}}]{Narayan2012MNRAS.426.3241N}
{Narayan}, R., {S{\"A} dowski}, A., {Penna}, R.~F., \& {Kulkarni}, A.~K. 2012,
  \mnras, 426, 3241

\bibitem[{{Narayan} \& {Yi}(1994)}]{Narayan1994ADAF_ApJ...428L..13N}
{Narayan}, R., \& {Yi}, I. 1994, \apjl, 428, L13

\bibitem[{{Narayan} \&
  {Yi}(1995{\natexlab{a}})}]{Narayan1995ADAF_ApJ...444..231N}
---. 1995{\natexlab{a}}, \apj, 444, 231

\bibitem[{{Narayan} \& {Yi}(1995{\natexlab{b}})}]{Narayan1995ApJ...452..710N}
---. 1995{\natexlab{b}}, \apj, 452, 710

\bibitem[{{Narayan} {et~al.}(1995){Narayan}, {Yi}, \&
  {Mahadevan}}]{Narayan1995Natur.374..623N}
{Narayan}, R., {Yi}, I., \& {Mahadevan}, R. 1995, \nat, 374, 623

\bibitem[{{O'Dea}(1998)}]{1998PASP..110..493O}
{O'Dea}, C.~P. 1998, \pasp, 110, 493

\bibitem[{{Oka} {et~al.}(2017){Oka}, {Tsujimoto}, {Iwata}, {Nomura}, \&
  {Takekawa}}]{Oka2017NatAs...1..709O}
{Oka}, T., {Tsujimoto}, S., {Iwata}, Y., {Nomura}, M., \& {Takekawa}, S. 2017,
  Nature Astronomy, 1, 709

\bibitem[{{Orienti} \& {Dallacasa}(2008)}]{2008A&A...487..885O}
{Orienti}, M., \& {Dallacasa}, D. 2008, \aap, 487, 885

\bibitem[{{Paczy{\'n}sky} \& {Wiita}(1980)}]{Paczynsky1980A&A....88...23P}
{Paczy{\'n}sky}, B., \& {Wiita}, P.~J. 1980, \aap, 500, 203

\bibitem[{{Quataert} \&
  {Gruzinov}(2000)}]{Quataert&Gruzinov2000ApJ...539..809Q}
{Quataert}, E., \& {Gruzinov}, A. 2000, \apj, 539, 809

\bibitem[{{Ravi} {et~al.}(2018){Ravi}, {Vedantham}, \&
  {Phinney}}]{Ravi2018MNRAS.478L..72R}
{Ravi}, V., {Vedantham}, H., \& {Phinney}, E.~S. 2018, \mnras, 478, L72

\bibitem[{{Reines} {et~al.}(2020){Reines}, {Condon}, {Darling}, \&
  {Greene}}]{Reines2020ApJ...888...36R}
{Reines}, A.~E., {Condon}, J.~J., {Darling}, J., \& {Greene}, J.~E. 2020, \apj,
  888, 36

\bibitem[{{Ressler} {et~al.}(2018){Ressler}, {Quataert}, \&
  {Stone}}]{Ressler2018MNRAS.478.3544R}
{Ressler}, S.~M., {Quataert}, E., \& {Stone}, J.~M. 2018, \mnras, 478, 3544

\bibitem[{{Ressler} {et~al.}(2020){Ressler}, {Quataert}, \&
  {Stone}}]{Ressler2020MNRAS.492.3272R}
---. 2020, MNRAS, 492, 3272

\bibitem[{{Ruffert} \& {Arnett}(1994)}]{Ruffert1994ApJ...427..351R}
{Ruffert}, M., \& {Arnett}, D. 1994, \apj, 427, 351

\bibitem[{{Russell} {et~al.}(2015){Russell}, {Fabian}, {McNamara}, \&
  {Broderick}}]{Russell2015MNRAS.451..588R}
{Russell}, H.~R., {Fabian}, A.~C., {McNamara}, B.~R., \& {Broderick}, A.~E.
  2015, \mnras, 451, 588

\bibitem[{{Russell} {et~al.}(2013){Russell}, {McNamara}, {Edge}, {Hogan},
  {Main}, \& {Vantyghem}}]{Russell2013MNRAS.432..530R}
{Russell}, H.~R., {McNamara}, B.~R., {Edge}, A.~C., {et~al.} 2013, \mnras, 432,
  530

\bibitem[{{Ryan} {et~al.}(2017){Ryan}, {Ressler}, {Dolence}, {Tchekhovskoy},
  {Gammie}, \& {Quataert}}]{Ryan2017ApJ...844L..24R}
{Ryan}, B.~R., {Ressler}, S.~M., {Dolence}, J.~C., {et~al.} 2017, \apjl, 844,
  L24

\bibitem[{{Ryu} {et~al.}(2018){Ryu}, {Perna}, {Haiman}, {Ostriker}, \&
  {Stone}}]{Ryu2018MNRAS.473.3410R}
{Ryu}, T., {Perna}, R., {Haiman}, Z., {Ostriker}, J.~P., \& {Stone}, N.~C.
  2018, \mnras, 473, 3410

\bibitem[{{Sadowski} {et~al.}(2015){Sadowski}, {Narayan}, {Tchekhovskoy},
  {Abarca}, {Zhu}, \& {McKinney}}]{Sadowski2015MNRAS.447...49S}
{Sadowski}, A., {Narayan}, R., {Tchekhovskoy}, A., {et~al.} 2015, \mnras, 447,
  49

\bibitem[{{Schneider} {et~al.}(2002){Schneider}, {Ferrara}, {Natarajan}, \&
  {Omukai}}]{Schneider2002ApJ...571...30S}
{Schneider}, R., {Ferrara}, A., {Natarajan}, P., \& {Omukai}, K. 2002, \apj,
  571, 30

\bibitem[{{Schulze} \& {Wisotzki}(2010)}]{Schulze2010A&A...516A..87S}
{Schulze}, A., \& {Wisotzki}, L. 2010, \aap, 516, A87

\bibitem[{{Sesana} {et~al.}(2008){Sesana}, {Vecchio}, \&
  {Colacino}}]{Sesana2008MNRAS.390..192S}
{Sesana}, A., {Vecchio}, A., \& {Colacino}, C.~N. 2008, \mnras, 390, 192

\bibitem[{{Shakura} \& {Sunyaev}(1973)}]{Shakura&Sunyaev1973A&A....24..337S}
{Shakura}, N.~I., \& {Sunyaev}, R.~A. 1973, \aap, 500, 33

\bibitem[{{Shima} {et~al.}(1985){Shima}, {Matsuda}, {Takeda}, \&
  {Sawada}}]{Shima1985MNRAS.217..367S}
{Shima}, E., {Matsuda}, T., {Takeda}, H., \& {Sawada}, K. 1985, \mnras, 217,
  367

\bibitem[{{Sikora} {et~al.}(2007){Sikora}, {Stawarz}, \&
  {Lasota}}]{Sikora2007ApJ...658..815S}
{Sikora}, M., {Stawarz}, {\L}., \& {Lasota}, J.-P. 2007, \apj, 658, 815

\bibitem[{{Stone} \& {Norman}(1992)}]{1992ApJS...80..753S}
{Stone}, J.~M., \& {Norman}, M.~L. 1992, \apjs, 80, 753

\bibitem[{{Stone} {et~al.}(2019){Stone}, {Tomida}, {White}, \&
  {Felker}}]{Athena2019ascl.soft12005S}
{Stone}, J.~M., {Tomida}, K., {White}, C., \& {Felker}, K.~G. 2019, {Athena++:
  Radiation GR magnetohydrodynamics code}, , , ascl:1912.005

\bibitem[{{Thomas} {et~al.}(2005){Thomas}, {Maraston}, {Bender}, \& {Mendes de
  Oliveira}}]{Thomas2005ApJ...621..673T}
{Thomas}, D., {Maraston}, C., {Bender}, R., \& {Mendes de Oliveira}, C. 2005,
  \apj, 621, 673

\bibitem[{{Thompson} {et~al.}(1980){Thompson}, {Clark}, {Wade}, \&
  {Napier}}]{VLA1980ApJS...44..151T}
{Thompson}, A.~R., {Clark}, B.~G., {Wade}, C.~M., \& {Napier}, P.~J. 1980,
  \apjs, 44, 151

\bibitem[{{Tremmel} {et~al.}(2018{\natexlab{a}}){Tremmel}, {Governato},
  {Volonteri}, {Pontzen}, \& {Quinn}}]{Tremmel2018ApJ...857L..22T}
{Tremmel}, M., {Governato}, F., {Volonteri}, M., {Pontzen}, A., \& {Quinn},
  T.~R. 2018{\natexlab{a}}, \apjl, 857, L22

\bibitem[{{Tremmel} {et~al.}(2018{\natexlab{b}}){Tremmel}, {Governato},
  {Volonteri}, {Quinn}, \& {Pontzen}}]{Tremmel2018MNRAS.475.4967T}
{Tremmel}, M., {Governato}, F., {Volonteri}, M., {Quinn}, T.~R., \& {Pontzen},
  A. 2018{\natexlab{b}}, \mnras, 475, 4967

\bibitem[{{Tsuboi} {et~al.}(2017){Tsuboi}, {Kitamura}, {Tsutsumi}, {Uehara},
  {Miyoshi}, {Miyawaki}, \& {Miyazaki}}]{Tsuboi2017ApJ...850L...5T}
{Tsuboi}, M., {Kitamura}, Y., {Tsutsumi}, T., {et~al.} 2017, \apjl, 850, L5

\bibitem[{{Tsuna} {et~al.}(2018){Tsuna}, {Kawanaka}, \&
  {Totani}}]{Tsuna2018MNRAS.477..791T}
{Tsuna}, D., {Kawanaka}, N., \& {Totani}, T. 2018, \mnras, 477, 791

\bibitem[{{Volonteri} \& {Perna}(2005)}]{Volonteri2005MNRAS.358..913V}
{Volonteri}, M., \& {Perna}, R. 2005, \mnras, 358, 913

\bibitem[{{Xie} \& {Yuan}(2012)}]{Xie2012MNRAS.427.1580X}
{Xie}, F.-G., \& {Yuan}, F. 2012, \mnras, 427, 1580

\bibitem[{{Xu} \& {Stone}(2019)}]{Xu2019MNRAS.488.5162X}
{Xu}, W., \& {Stone}, J.~M. 2019, \mnras, 488, 5162

\bibitem[{{Yu}(2002)}]{Yu2002MNRAS.331..935Y}
{Yu}, Q. 2002, \mnras, 331, 935

\bibitem[{{Yuan} \& {Narayan}(2014)}]{Yuan2014ARA&A..52..529Y}
{Yuan}, F., \& {Narayan}, R. 2014, \araa, 52, 529

\bibitem[{{Yuan} {et~al.}(2000){Yuan}, {Peng}, {Lu}, \&
  {Wang}}]{Yuan2000ApJ...537..236Y}
{Yuan}, F., {Peng}, Q., {Lu}, J.-f., \& {Wang}, J. 2000, \apj, 537, 236

\bibitem[{{Yuan} {et~al.}(2004){Yuan}, {Quataert}, \&
  {Narayan}}]{Yuan2004ApJ...606..894Y}
{Yuan}, F., {Quataert}, E., \& {Narayan}, R. 2004, \apj, 606, 894

\bibitem[{{Yuan} {et~al.}(2012){Yuan}, {Wu}, \& {Bu}}]{Yuan2012ApJ...761..129Y}
{Yuan}, F., {Wu}, M., \& {Bu}, D. 2012, \apj, 761, 129

\bibitem[{{Zivancev} {et~al.}(2020){Zivancev}, {Ostriker}, \&
  {Kupper}}]{Zivancev&Ostriker2020arXiv200406083Z}
{Zivancev}, C., {Ostriker}, J., \& {Kupper}, A. H.~W. 2020, arXiv e-prints,
  arXiv:2004.06083

\end{thebibliography}

\end{document}